\pdfoutput=1

\documentclass[11pt]{article}

\usepackage{ACL2023}

\usepackage{times}
\usepackage{latexsym}

\usepackage[T1]{fontenc}

\usepackage[utf8]{inputenc}

\usepackage{microtype}

\usepackage{inconsolata}

\usepackage{graphicx}
\usepackage{booktabs}
\usepackage{multirow}

\title{Inspecting Spoken Language Understanding from Kids for Basic~Math~Learning~at~Home}

\author{Eda Okur \\ Intel Labs, USA \\ \texttt{eda.okur@intel.com}
        \And Roddy Fuentes Alba \\ Intel Labs, Mexico \\ \texttt{roddy.fuentes.alba@intel.com}
        \AND Saurav Sahay \\ Intel Labs, USA \\ \texttt{saurav.sahay@intel.com}
        \And Lama Nachman \\ Intel Labs, USA \\ \texttt{lama.nachman@intel.com}
        \\}

\begin{document}
\maketitle
\begin{abstract}

Enriching the quality of early childhood education with interactive math learning at home systems, empowered by recent advances in conversational AI technologies, is slowly becoming a reality. With this motivation, we implement a multimodal dialogue system to support play-based learning experiences at home, guiding kids to master basic math concepts. This work explores Spoken Language Understanding (SLU) pipeline within a task-oriented dialogue system developed for Kid Space, with cascading Automatic Speech Recognition (ASR) and Natural Language Understanding (NLU) components evaluated on our home deployment data with kids going through gamified math learning activities. We validate the advantages of a multi-task architecture for NLU and experiment with a diverse set of pretrained language representations for Intent Recognition and Entity Extraction tasks in the math learning domain. To recognize kids' speech in realistic home environments, we investigate several ASR systems, including the commercial Google Cloud and the latest open-source Whisper solutions with varying model sizes. We evaluate the SLU pipeline by testing our best-performing NLU models on noisy ASR output to inspect the challenges of understanding children for math learning in authentic homes.

\end{abstract}

\section{Introduction and Background}

The ongoing progress in Artificial Intelligence (AI) based advanced technologies can assist humanity in reducing the most critical inequities around the globe. The recent widespread interest in conversational AI applications presents exciting opportunities to showcase the positive societal impact of these technologies. The language-based AI systems have already started to mature to a level where we may soon observe their influences in mitigating the most pressing global challenges. Education is among the top priority improvement areas identified by the United Nations (UN) (i.e., poverty, hunger, healthcare, and education). In particular, increasing the inclusiveness and quality of education is within the UN development goals\footnote{\url{https://sdgs.un.org/goals}} with utmost urgency. One of the preeminent ways to diminish societal inequity is promoting STEM (i.e., Science, Technology, Engineering, Math) education, specifically ensuring that children succeed in mathematics. It is well-known that acquiring basic math skills at younger ages builds students up for success, regardless of their future career choices~\cite{cesarone2008early, torpey2012math}. For math education, interactive learning environments through gamification present substantial leverages over more traditional learning settings for studying elementary math subjects, particularly with younger learners~\cite{skene2022can}. With that goal, conversational AI technologies can facilitate this interactive learning environment where students can master fundamental math concepts. Despite these motivations, studying spoken language technologies for younger kids to learn basic math is a vastly uncharted area of AI.

This work discusses a modular goal-oriented Spoken Dialogue System (SDS) specifically targeted for kids to learn and practice basic math concepts at home setup. Initially, a multimodal dialogue system~\cite{KidSpace-SemDial-2019} is implemented for Kid Space~\cite{KidSpace-ICMI-2018}, a gamified math learning application for deployment in authentic classrooms. During this preliminary real-world deployment at an elementary school, the COVID-19 pandemic impacted the globe, and school closures forced students to switch to online learning options at home. To support this sudden paradigm shift to at-home learning, previous school use cases are redesigned for new home usages, and our dialogue system is recreated to deal with interactive math games at home. While the play-based learning activities are adjusted for home usages with a much simpler setup, the multimodal aspects of these games are partially preserved along with the fundamental math concepts for early childhood education. These math skills cover using ones and tens to construct numbers and foundational arithmetic concepts and operations such as counting, addition, and subtraction. The multimodal aspects of these learning games include kids' spoken interactions with the system while answering math questions and carrying out game-related conversations, physical interactions with the objects (i.e., placing cubes and sticks as manipulatives) on a visually observed playmat, performing specific pose and gesture-based actions as part of these interactive games (e.g., jumping, standing still, air high-five).

\begin{figure}[t]
\begin{center} 
\includegraphics[width=\columnwidth]{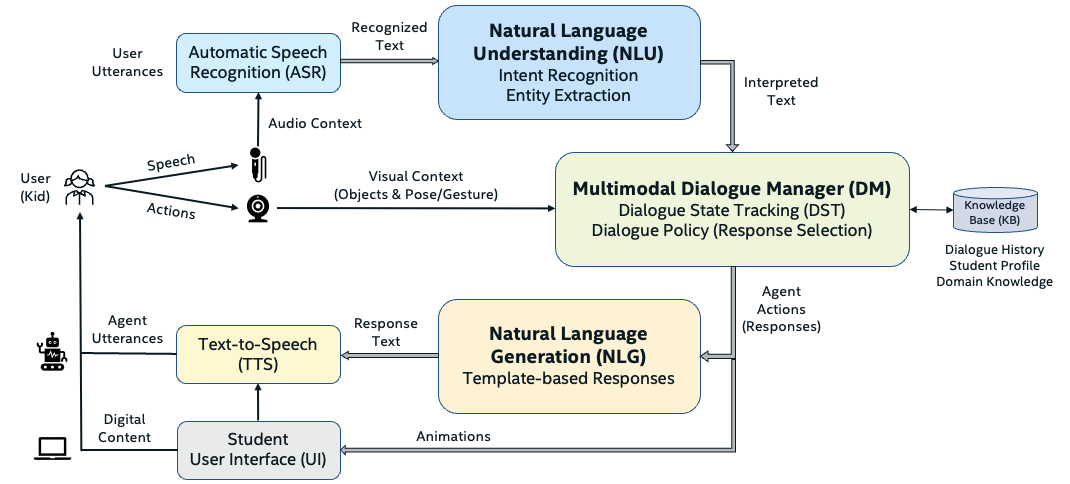}
\caption{Multimodal Dialogue System Pipeline}
\label{fig-pipeline}
\end{center}
\end{figure}

Our domain-specific SDS pipeline (see Figure~\ref{fig-pipeline}) consists of multiple cascaded components, namely Automatic Speech Recognition (ASR), Natural Language Understanding (NLU), Multimodal Dialogue Manager (DM), Natural Language Generation (NLG), and Text-to-Speech (TTS) synchronizing the agent utterances with virtual character animations on Student User Interface (UI). Here we concentrate on the Spoken Language Understanding (SLU) task on kids' speech at home environments while playing basic math games. Such application-dependent SLU approaches commonly involve two main modules applied sequentially: (i) Speech-to-Text (STT) or ASR module that recognizes speech and transcribes the spoken utterances into text, and (ii) NLU module that interprets the semantics of those utterances by processing the transcribed text. NLU is one of the most integral components of these goal-oriented dialogue systems. It empowers user-agent interactions by understanding the meaning of user utterances via performing domain-specific sub-tasks. Intent Recognition (IR) and Named Entity Recognition (NER) are essential sub-tasks within the NLU module to resolve the complexities of human language and extract meaningful information for the application at hand. Given a user utterance as input, the Intent Classification aims to identify the user's intention (i.e., what the user desires to achieve with that interaction) and categorize the user's objective at that conversational turn. The Entity Extraction targets locating and classifying entities (i.e., specific terms representing existing things such as person names, locations, and organizations) mentioned in user utterances into predefined task-specific categories.

In this study, we present our efforts to convert the task-oriented SDS~\cite{KidSpace-LREC-2022} designed for school use cases~\cite{KidSpace-ETRD-2022} to home usages after COVID-19 and inspect the performance of individual SDS modules evaluated on the home deployment data we recently collected from 12 kids individually at their homes. The current work focuses on assessing and improving the SLU task performance on kids' utterances at home by utilizing this real-world deployment data. We first investigate the ASR and NLU module evaluations independently. Then, we inspect the overall SLU pipeline (ASR+NLU) performance on kids' speech by evaluating our NLU tasks on ASR output (i.e., recognized text) at home environments. As the erroneous and noisy speech recognition output would lead to incorrect intent and entity predictions, we aim to understand these error propagation consequences with SLU for children in the math learning domain. We experiment with various recent ASR solutions and diverse model sizes to gain more insights into their capabilities to recognize kids' speech at home. We then analyze the effects of these ASR engines on understanding intents and extracting entities from children's utterances. We discuss our findings and observations for potential enhancements in future deployments of this multimodal dialogue system for math learning at home.

\section{Related Work}

\subsection{Conversational AI for Math Learning}

With the ultimate goal of improving the quality of education, there has been a growing enthusiasm for exploiting AI-based intelligent systems to boost students' learning experiences~\cite{CHASSIGNOL201816,AdaptLE-CHI-2019,MMHCI,zhai2021review,baker2021artificial}. Among these, interactive frameworks that support guided play-based learning spaces revealed significant advantages for math learning~\cite{10.3389/feduc.2019.00081,sungamifying,10.1007/978-3-030-78292-4_28}, especially for building foundational math skills in early childhood education~\cite{kumarphygital,skene2022can}. To attain this level of interactivity within smart learning spaces, developing innovative educational applications by utilizing language-based AI technologies is in growing demand~\cite{taghipour-ng-2016-neural,lende2016question,raamadhurai-etal-2019-curio,cahill-etal-2020-context,DBLP:journals/corr/abs-2112-01012,rathod-etal-2022-educational}. In particular, designing conversational agents for intelligent tutoring is a compelling yet challenging area of research, with several attempts presented so far~\cite{alex254848,wambsganss2020conversational,winkler2020sara,DBLP:journals/corr/abs-2010-12710,okonkwo2021chatbots,wollny2021we}, most of them focusing on language learning~\cite{bibauw2022dialogue,tyen-etal-2022-towards,zhang2022storybuddy}. 

In the math education context, earlier conversational math tutoring applications exist, such as SKOPE-IT~\cite{nye2018skope}, which is based on AutoTutor~\cite{1532370} and ALEKS~\cite{falmagne2013knowledge}, and MathBot~\cite{grossman2019mathbot}. These are often text-based online systems following strict rules in conversational graphs. Later, various studies emerged at the intersection of cutting-edge AI techniques and math learning~\cite{mansouri2019tangent,huangreal2,azerbayev2022proofnet,uesatosolving,yang-etal-2022-logicsolver}. Among those, employing advanced language understanding methods to assist math learning is relatively new~\cite{DBLP:journals/corr/abs-2105-00377,MathBERT-2021,loginova2022structural,reusch2022transformer}. The majority of those recent work leans on exploring language representations for math-related tasks such as mathematical reasoning, formula understanding, math word problem-solving, knowledge tracing, and auto-grading, to name a few. Recently, TalkMoves dataset~\cite{suresh-etal-2022-talkmoves} was released with K-12 math lesson transcripts annotated for discursive moves and dialogue acts to classify teacher talk moves in math classrooms~\cite{suresh-etal-2022-fine}.

For the conversational AI tasks, the latest large language models (LLMs) based chatbots, such as BlenderBot~\cite{shuster2022blenderbot} and ChatGPT~\cite{team2022chatgpt}, gained a lot of traction in the education community~\cite{tack2022ai,kasneci2023chatgpt}, along with some concerns about using generative models in tutoring~\cite{macina2023opportunities,cotton2023chatting}. ChatGPT is a general-purpose open-ended interaction agent trained on internet-scale data. It is an end-to-end dialogue model without explicit NLU/Intent Recognizer or DM, which currently cannot fully comprehend the multimodal context and proactively generate responses to nudge children in a guided manner without distractions. Using these recent chatbots for math learning is still in the early stages because they are known to miss basic mathematical abilities and carry reasoning flaws~\cite{frieder2023mathematical}, revealing a lack of common sense. Moreover, they are known to be susceptible to triggering inappropriate or harmful responses and potentially perpetuate human biases since they are trained on internet-scale data and require carefully-thought guardrails.

On the contrary, our unique application is a task-oriented math learning spoken dialogue system designed to perform learning activities, following structured educational games to assist kids in practicing basic math concepts at home. Our SDS does not require massive amounts of data to understand kids and generate appropriate adaptive responses, and the lightweight models can run locally on client machines. In addition, our solution is multimodal, intermixing the physical and digital hybrid learning experience with audio-visual understanding, object recognition, segmentation, tracking, and pose and gesture recognition.





\subsection{Spoken Language Understanding}

Conventional pipeline-based dialogue systems with supervised learning are broadly favored when initial domain-specific training data is scarce to bootstrap the task-oriented SDS for future data collection~\cite{serban2018survey,budzianowski-etal-2018-multiwoz,DBLP:journals/corr/abs-2009-13570}. Deep learning-based modular dialogue frameworks and practical toolkits are prominent in academic and industrial settings~\cite{bocklisch2017rasa,burtsev-etal-2018-deeppavlov,reyes2019methodology}. For task-specific applications with limited in-domain data, current SLU systems often use a cascade of two neural modules: (i) ASR maps the input audio to text (i.e., transcript), and (ii) NLU predicts intent and slots/entities from this transcript. Since our main focus in this work is investigating the SLU pipeline, we briefly summarize the existing NLU and ASR solutions.

\subsubsection{Language Representations for NLU}
\label{sec:lang-reps}

The NLU component processes input text, often detects intents, and extracts referred entities from user utterances. For the mainstream NLU tasks of Intent Classification and Entity Recognition, jointly trained multi-task models are proposed~\cite{Liu+2016,zhang-2016,goo-etal-2018-slot} with hierarchical learning approaches~\cite{wen-2018,AMIE-CICLing-2019,vanzo-etal-2019-hierarchical}. Transformer architecture~\cite{DBLP:conf/nips/VaswaniSPUJGKP17} is a game-changer for several downstream language tasks. With Transformers, BERT~\cite{devlin-etal-2019-bert} is presented, which became one of the most pivotal breakthroughs in language representations, achieving high performance in various tasks, including NLU. Later, Dual Intent and Entity Transformer (DIET) architecture~\cite{DIET-2020} is invented as a lightweight multi-task NLU model. On multi-domain NLU-Benchmark data~\cite{liu2021benchmarking}, the DIET model outperformed fine-tuning BERT for joint Intent and Entity Recognition.

For BERT-based autoencoding approaches, RoBERTa~\cite{liu2019roberta} is presented as a robustly optimized BERT model for sequence and token classification. The Hugging Face introduced a smaller, lighter general-purpose language representation model called DistilBERT~\cite{sanh2019distilbert} as the knowledge-distilled version of BERT. ConveRT~\cite{ConveRT-2020} is proposed as an efficiently compact model to obtain pretrained sentence embeddings as conversational representations for dialogue-specific tasks. LaBSE~\cite{feng-etal-2022-language} is a pretrained multilingual model producing language-agnostic BERT sentence embeddings that achieve promising results in text classification. 

The GPT family of autoregressive LLMs, such as GPT-2~\cite{radford2019language} and GPT-3~\cite{brown2020language}, perform well at what they are pretrained for, i.e., text generation. GPT models can also be adopted for NLU, supporting few-shot learning capabilities, and NLG in task-oriented dialogue systems~\cite{madotto2020language,liu2021gpt}. XLNet~\cite{yang2019xlnet} applies autoregressive pretraining for representation learning that adopts Transformer-XL~\cite{dai-etal-2019-transformer} as a backbone model and works well for language tasks with lengthy contexts. DialoGPT~\cite{zhang-etal-2020-dialogpt} extends GPT-2 as a large-scale neural response generation model for multi-turn conversations trained on Reddit discussions, whose representations can be exploited in dialogue tasks.

For language representations to be utilized in math-related tasks, MathBERT~\cite{MathBERT-2021} is introduced as a math-specific BERT model pretrained on large math corpora. Later, Math-aware-BERT and Math-aware-RoBERTa models~\cite{reusch2022transformer} are proposed based on BERT and RoBERTa, pretrained on Math Stack Exchange\footnote{\url{https://math.stackexchange.com/}}. 


\subsubsection{Speech Recognition with Kids}
\label{sec:speech-rec}

Speech recognition technology has been around for some time, and numerous ASR solutions are available today, both commercial and open-source. Rockhopper ASR~\cite{stemmer2017speech} is an earlier low-power speech recognition engine with LSTM-based language models, where its acoustic models are trained using an open-source Kaldi speech recognition toolkit~\cite{povey2011kaldi}. Google Cloud Speech-to-Text\footnote{\url{https://cloud.google.com/speech-to-text/}} is a prominent commercial ASR service powered by advanced neural models and designed for speech-dependant applications. Until recently, Google STT API was arguably the leader in ASR services for recognition performance and language coverage. \newcite{ASRbenchmark2018} reported that Google ASR could reach a word error rate (WER) of 12.1\% on LibriSpeech clean dataset (28.8\% on LibriSpeech other)~\cite{panayotov2015librispeech} at that time, which is improved drastically over time. Recently, Open AI released Whisper ASR~\cite{radford2022robust} as a game-changer speech recognizer. Whisper models are pretrained on a vast amount of labeled audio-transcription data (i.e., 680k hours), unlike its predecessors (e.g., Wav2Vec 2.0~\cite{baevski2020wav2vec} is trained on 60k hours of unlabeled audio). 117k hours of this data are multilingual, which makes Whisper applicable to over 96 languages, including low-resourced ones. Whisper architecture follows a standard Transformer-based encoder-decoder as many speech-related models~\cite{latif2023transformers}. The Whisper-base model is reported to achieve 5.0\% \& 12.4\% WER on LibriSpeech clean \& other.

Although speech recognition systems are substantially improving to achieve human recognition levels, problems still occur, especially in noisy environments, with users having accents and dialects or underrepresented groups like kids. Child speech brings distinct challenges to ASR~\cite{stemmer2003acoustic,gerosa2007acoustic,yeung2018difficulties}, such as data scarcity and highly varied acoustic, linguistic, physiological, developmental, and articulatory characteristics compared to adult speech~\cite{claus2013survey,shivakumar2020transfer,bhardwaj2022automatic}. Thus, WER for children's voices is reported two-to-five times worse than for adults~\cite{wu2019advances}, as the younger the child, the poorer ASR performs. There exist efforts to mitigate these difficulties of speech recognition with kids~\cite{shivakumar2014,duan2020unsupervised,booth-etal-2020-evaluating,kelly2020soapbox,rumberg2021age,yeung2021fundamental}. Few studies also focus on speech technologies in educational settings~\cite{reeder2015speech,10.1007/978-3-319-19773-9_3,Bai2021,bai22b_interspeech,dutta-etal-2022-activity}, often for language acquisition, reading comprehension, and story-telling activities.

\section{Methods}

\subsection{Home Learning Data and Use Cases}
\label{sec:data}

We utilize two datasets for gamified basic math learning at home usages. The first set is a proof-of-concept (POC) data manually constructed based on User Experience (UX) studies (e.g., detailed scripts for new home use cases) and partially adopted from our previous school data~\cite{KidSpace-MathNLP-EMNLP-2022}. This POC data is used to train and cross-validate various NLU models to develop the best practices in later home deployments. The second set is our recent home deployment data collected from 12 kids (ages 7-8) experiencing our multimodal math learning system at authentic homes. The audio-visual data is transcribed manually, and user utterances in these reference transcripts are annotated for intent and entity types we identified for each learning activity at home. Table~\ref{data-stats} compares the NLU statistics for Kid Space Home POC and Deployment datasets. Manually transcribed children's utterances in deployment data are employed to test our best NLU models trained on POC data. We run multiple ASR engines on audio recordings from home deployment data, where automatic transcripts (i.e., ASR output) are utilized to compute WER to assess ASR model performances on kids' speech. We also evaluate the SLU pipeline (ASR+NLU) by testing NLU models on ASR output from deployment data.

The simplified home deployment setup includes a playmat with physical manipulatives, a laptop with a built-in camera, a wireless lavalier mic, and a depth camera on a tripod. Home use cases follow a particular flow of activities designed for play-based learning in early childhood education. These activities are Introduction (Meet \& Greet), Warm-up Game (Red Light Green Light), Training Game, Learning Game, and Closure (Dance Party). After meeting with the virtual character and playing jumping games, the child starts the training game, where the agent asks for help planting flowers. The agent presents tangible manipulatives, cubes representing ones and sticks representing tens, and instructs the kid to answer basic math questions and construct numbers using these objects, going through multiple rounds of practice questions where flowers in child-selected colors bloom as rewards. In the actual learning game, the agent presents clusters of questions involving ones \& tens, and the child provides verbal (e.g., stating the numbers) and visual answers (e.g., placing the cubes and sticks on the playmat, detected by the overhead camera). The agent provides scaffolding utterances and performs animations to show and tell how to solve basic math questions. The interaction ends with a dance party to celebrate achievements and say goodbyes in closure. Some of our intents can be considered generic (e.g., \textit{state-name}, \textit{affirm}, \textit{deny}, \textit{repeat}, \textit{out-of-scope}), but some are highly domain-specific (e.g., \textit{answer-flowers}, \textit{answer-valid}, \textit{answer-others}, \textit{state-color}, \textit{had-fun-a-lot}, \textit{end-game}) or math-related (e.g., \textit{state-number}, \textit{still-counting}). The entities we extract are activity-specific (i.e., \textit{name}, \textit{color}) and math-related (i.e., \textit{number}).

\begin{table}[!t]
  \centering
  \small
  \resizebox{\columnwidth}{!}{
  \begin{tabular}{lcc}
    \toprule
    \textbf{NLU Data Statistics} & \textbf{POC} & \textbf{Deployment} \\
    \midrule
    \# Intents Types & 13 & 12 \\
    Total \# Utterances & 4091 & 733 \\
    \midrule
    \# Entity Types & 3 & 3 \\
    Total \# Entities & 2244 & 497 \\
    \midrule
    Min \# Utterances per Intent & 105 & 1 \\
    Max \# Utterances per Intent & 830 & 270 \\
    Avg \# Utterances per Intent & 314.7 & 61.1 \\
    \midrule
    Min \# Tokens per Utterance & 1 & 1 \\
    Max \# Tokens per Utterance & 40 & 33 \\
    Avg \# Tokens per Utterance & 4.49 & 2.30 \\
    \midrule
    \# Unique Tokens (Vocab Size) & 702 & 149 \\
    Total \# Tokens & 18364 & 1689 \\
    \bottomrule
  \end{tabular}
  }
  \caption{Kid Space Home POC and Deployment Data}
  \label{data-stats}
\end{table}

\subsection{NLU and ASR Models}

Customizing open-source Rasa framework~\cite{bocklisch2017rasa} as a backbone, we investigate several NLU models for Intent Recognition and Entity Extraction tasks to implement our math learning conversational AI system for home usage. Our baseline approach is inspired by the StarSpace~\cite{wu2018starspace} method, a supervised embedding-based model maximizing the similarity between utterances and intents in shared vector space. We enrich this simple text classifier by incorporating SpaCy~\cite{honnibal10boyd} pretrained language models\footnote{\url{https://github.com/explosion/spacy-models/releases/tag/en_core_web_md-3.5.0}} for word embeddings as additional features in the NLU pipeline. CRF Entity Extractor~\cite{DBLP:conf/icml/LaffertyMP01} with BILOU tagging is also part of this baseline NLU. For home usages, we explore the advantages of switching to a more recent DIET model\footnote{Please check~\newcite{DIET-2020} for hyper-parameter tuning, hardware specs, and computational costs.} for joint Intent and Entity Recognition, a multi-task architecture with two-layer Transformers shared for NLU tasks. DIET leverages combining dense features (e.g., any given pretrained embeddings) with sparse features (e.g., token-level encodings of char n-grams). To observe the net benefits of DIET, we first pass the identical SpaCy embeddings used in our baseline (StarSpace) as dense features to DIET. Then, we adopt DIET with pretrained BERT\footnote{\url{https://huggingface.co/bert-base-uncased}}, RoBERTa\footnote{\url{https://huggingface.co/roberta-base}}, and DistilBERT\footnote{\url{https://huggingface.co/distilbert-base-uncased}} word embeddings, as well as ConveRT\footnote{\url{https://github.com/connorbrinton/polyai-models/releases}} and LaBSE\footnote{\url{https://huggingface.co/rasa/LaBSE}} sentence embeddings to inspect the effects of these autoencoding-based language representations on NLU performance (see~\ref{sec:lang-reps} for more details). We also evaluate pretrained embeddings from models using autoregressive training such as XLNet\footnote{\url{https://huggingface.co/xlnet-base-cased}}, GPT-2\footnote{\url{https://huggingface.co/gpt2}}\footnote{Excluded GPT-3 and beyond that are not open-source.}, and DialoGPT\footnote{\url{https://huggingface.co/microsoft/DialoGPT-medium}} on top of DIET. Next, we explore recently-proposed math-language representations pretrained on math data for our basic math learning dialogue system. MathBERT~\cite{MathBERT-2021} is pretrained on large math corpora (e.g., curriculum, textbooks, MOOCs, arXiv papers) covering pre-k to college-graduate materials. We enhance DIET by incorporating embeddings from MathBERT-base\footnote{\url{https://huggingface.co/tbs17/MathBERT}} and MathBERT-custom\footnote{\url{https://huggingface.co/tbs17/MathBERT-custom}} models, pretrained with BERT-base original and math-customized vocabularies, respectively. Math-aware-BERT\footnote{\url{https://huggingface.co/AnReu/math_pretrained_bert}} and Math-aware-RoBERTa\footnote{\url{https://huggingface.co/AnReu/math_pretrained_roberta}} models~\cite{reusch2022transformer} are initialized from BERT-base and RoBERTa-base, and further pretrained on Math StackExchange\footnote{\url{https://archive.org/download/stackexchange}} with extra LaTeX tokens to better tokenize math formulas for ARQMath-3 tasks~\cite{10.1007/978-3-031-13643-6_20}. We exploit these representations with DIET to investigate their effects on our NLU tasks in the basic math domain.

For the ASR module, we explore three main speech recognizers for our math learning application at home, which are explained further in~\ref{sec:speech-rec}. Rockhopper ASR\footnote{\url{https://docs.openvino.ai/2018_R5/_samples_speech_sample_README.html}} is the baseline local approach previously inspected, which can be adjusted slightly for kids. Its acoustic models rely on Kaldi\footnote{\url{https://github.com/kaldi-asr/kaldi}} generated resources and are trained on default adult speech data. In the past explorations, when Rockhopper's language models fine-tuned with limited in-domain kids' utterances~\cite{KS2.0-DaSH-NAACL-2021} from previous school usages, WER decreased by 40\% for kids but remained 50\% higher than adult WER. Although this small-scale baseline solution is unexpected to reach Google Cloud ASR performance, Rockhopper has a few other advantages for our application since it can run offline locally on low-power devices, which could be better for security, privacy, latency, and cost (relative to cloud-based ASR services). Google ASR is a commercial cloud solution providing high-quality speech recognition service but requiring connectivity and payment, which cannot be adapted or fine-tuned as Rockhopper. The third ASR approach we investigate is Whisper\footnote{\url{https://github.com/openai/whisper}}, which combines the best of both worlds as it is an open-source adjustable solution that can run locally, achieving new state-of-the-art (SOTA) results. We inspect three configurations of varying model sizes (i.e., base, small, and medium) to evaluate the Whisper ASR for our home math learning usage with kids.

\section{Experimental Results}

To build the NLU module of our SLU pipeline, we train Intent and Entity Classification models and cross-validate them over the Kid Space Home POC dataset to decide upon the best-performing NLU architectures moving forward for home. Table~\ref{nlu-results-poc} summarizes the results of model selection experiments with various NLU models. We report the average of 5 runs, and each run involves a 10-fold cross-validation (CV) on POC data. Compared to the baseline StarSpace algorithm, we gain almost 2\% F1 score for intents and more than 1\% F1 for entities with multi-task DIET architecture. For language representations, we observe that incorporating DIET with the BERT family of embeddings from autoencoders achieves higher F1 scores relative to the GPT family of embeddings from autoregressive models. We cannot reveal any benefits of employing math-specific representations with DIET, as all such models achieve worse than DIET+BERT results. One reason we identify is the mismatch between our early math domain and advanced math corpora, including college-level math symbols and equations, that these models trained on. Another reason could be that such embeddings are pretrained on smaller math corpora (e.g., 100 million tokens) compared to massive-scale generic corpora (e.g., 3.3 billion words) that BERT models use for training. DIET+ConveRT is the clear winner for intents and achieves second-best but very close results for entities compared to DIET+LaBSE. ConveRT and LaBSE are both sentence-level embeddings, but ConveRT performs well on dialogue tasks as it is pretrained on large conversational corpora, including Reddit discussions. Based on these results, we select DIET+ConveRT as the final multi-task architecture for our NLU tasks at home.

\begin{table}[!t]
  \centering
  \small
  \resizebox{\columnwidth}{!}{
  \begin{tabular}{lcc}
    \toprule
    \textbf{NLU Model} & \textbf{Intent Detection} & \textbf{Entity Extraction} \\
    \midrule
    StarSpace+SpaCy & 92.71$\pm$0.25 & 97.08$\pm$0.21 \\
    DIET+SpaCy & 94.29$\pm$0.05 & 98.38$\pm$0.12 \\
    \midrule
    DIET+BERT & 97.25$\pm$0.23 & 99.23$\pm$0.02 \\
    DIET+RoBERTa & 95.50$\pm$0.18 & 99.11$\pm$0.12 \\
    DIET+DistilBERT & 97.41$\pm$0.20 & 99.49$\pm$0.12 \\
    DIET+ConveRT & \textbf{98.80}$\pm$\textbf{0.25} & 99.61$\pm$0.03 \\
    DIET+LaBSE & 98.19$\pm$0.18 & \textbf{99.72}$\pm$\textbf{0.04} \\
    \midrule
    DIET+XLNet & 94.99$\pm$0.19 & 98.38$\pm$0.14 \\
    DIET+GPT-2 & 95.35$\pm$0.27 & 99.01$\pm$0.27 \\
    DIET+DialoGPT & 96.00$\pm$0.49 & 98.94$\pm$0.12 \\
    \midrule
    DIET+MathBERT-base & 94.55$\pm$0.22 & 98.10$\pm$0.21 \\
    DIET+MathBERT-custom & 94.61$\pm$0.34 & 97.48$\pm$0.29 \\
    DIET+Math-aware-BERT & 95.95$\pm$0.15 & 98.94$\pm$0.19 \\
    DIET+Math-aware-RoBERTa & 94.20$\pm$0.16 & 98.75$\pm$0.21 \\
    \bottomrule
  \end{tabular}
  }
  \caption{NLU Model Selection Results in F1-scores (\%) Evaluated on Kid Space Home POC Data (10-fold CV)}
  \label{nlu-results-poc}
\end{table}

Next, we evaluate our NLU module on Kid Space Home Deployment data collected at authentic homes over 12 sessions with 12 kids. Each child goes through 5 activities within a session, as described in~\ref{sec:data}. In Table~\ref{nlu-results-dep}, we observe overall F1\% drops ($\Delta$) of 4.6 for intents and 0.3 for entities when our best-performing DIET+ConveRT models are tested on home deployment data. These findings are expected and relatively lower than the performance drops we previously observed at school~\cite{KidSpace-GAMNLP-LREC-2022}. We witness distributional and utterance-length differences between POC/training and deployment/test datasets. Real-world data would always be noisier than anticipated as these utterances come from younger kids playing math games in dynamic conditions.

To further improve the performance of our Kid Space Home NLU models (trained on POC data) by leveraging this recent deployment data, we experiment with merging the two datasets for training and evaluating the performance on individual deployment sessions via leave-one-out (LOO) CV. At each of the 12 runs (for 12 sessions/kids), we merge the POC data with 11 sessions of deployment data for model training and use the remaining session as a test set, then take the average performance of these runs. That would simulate how combining POC with real-world deployment data would help us train more robust NLU models that perform better on unseen data in future deployment sessions. The overall F1-scores reach 96.5\% for intents (2.3\% gain from 94.2\%) and 99.4\% for entities (0.1\% gain) with LOOCV, which are promising for our future deployments.

\begin{table}[!t]
  \centering
  \resizebox{\columnwidth}{!}{
  \begin{tabular}{lcccccc}
    \toprule
     & \multicolumn{3}{c}{\textbf{Intent Detection}} & \multicolumn{3}{c}{\textbf{Entity Extraction}}\\
    \textbf{Activity} & POC & Deploy & $\Delta$ & POC & Deploy & $\Delta$ \\
    \midrule
    Intro (Meet \& Greet) & 99.9 & 97.3 & -2.6 & 99.2 & 97.4 & -1.8 \\
    Warm-up Game & 98.8 & 93.4 & -5.4 & - & - & - \\
    Training Game & 98.4 & 94.2 & -4.2 & 99.9 & 99.8 & -0.1 \\
    Learning Game & 98.9 & 94.3 & -4.6 & 99.8 & 99.4 & -0.4 \\
    Closure (Dance) & 98.8 & 98.7 & -0.1 & - & - & - \\
    \midrule
    \textbf{All Activities} & \textbf{98.8} & \textbf{94.2} & \textbf{-4.6} & \textbf{99.6} & \textbf{99.3} & \textbf{-0.3} \\
    \bottomrule
  \end{tabular}
  }
  \caption{NLU Evaluation Results in F1-scores (\%) for DIET+ConveRT Models Trained on Kid Space Home POC Data \& Tested on Home Deployment Data}
  \label{nlu-results-dep}
\end{table}

\begin{table*}[t!]
  \centering
  \tiny
  \resizebox{\textwidth}{!}{
  \begin{tabular}{lcccccccc}
    \toprule
    \textbf{} & \textbf{Raw} & \textbf{Lowercase} & \textbf{Remove} & \textbf{Num2Word} & \textbf{LC \&} & \textbf{LC \& RP} & \textbf{NW \&} & \textbf{LC \& RP \&}\\
    \textbf{ASR Model} & \textbf{Output} & \textbf{(LC)} & \textbf{Punct (RP)} & \textbf{(NW)} & \textbf{RP} & \textbf{\& NW} & \textbf{Clean} & \textbf{NW \& Clean} \\
    \midrule
    Rockhopper & 0.939 & 0.919 & 0.924 & 0.937 & 0.886 & 0.884 & 0.937 & 0.884 \\
    Google Cloud & 0.829 & 0.798 & 0.775 & 0.763 & 0.695 & 0.602 & 0.763 & 0.602 \\
    Whisper-base & 1.042 & 1.020 & 0.971 & 0.985 & 0.946 & 0.856 & 0.622 & \textbf{0.500} \\
    Whisper-small & 0.834 & 0.804 & 0.760 & 0.756 & 0.720 & 0.621 & 0.537 & \textbf{0.405} \\
    Whisper-medium & 0.905 & 0.870 & 0.824 & 0.814 & 0.785 & 0.675 & 0.522 & \textbf{0.384} \\
    \bottomrule
  \end{tabular}
  }
  \caption{ASR Model Results: Avg Word Error Rates (WER) for Child Speech at Kid Space Home Deployment Data}
  \label{asr-wer-results}
\end{table*}

To inspect the ASR module of our SLU pipeline, we experiment with Rockhopper, Google, and Whisper-base/small/medium ASR models evaluated on the same audio data collected during home deployments. Using the manual session transcripts as a reference, we compute the average WER for kids with each ASR engine to investigate the most feasible solution. Table~\ref{asr-wer-results} summarizes WER results before and after standard pre-processing steps (e.g., lower casing and punctuation removal) as well as application-specific filters (e.g., num2word and cleaning). The numbers are transcribed inconsistently within reference transcripts plus ASR output (e.g., 35 vs. thirty-five), and we need to standardize them all in word forms. The cleaning step is applied to Whisper ASR output only due to known issues such as getting stuck in repeat loops and hallucinations~\cite{radford2022robust}. We seldom observe trash output from Whisper (4-to-7\%) having very long transcriptions with non-sense repetitions/symbols, which hugely affect WER due to their length, yet these samples can be easily auto-filtered. Even after these steps, the relatively high error rates can be attributed to many factors related to the characteristics of these recordings (e.g., incidental voice and phrases), very short utterances to be recognized (e.g., binary yes/no answers or stating numbers with one-or-two words), and recognizing kids' speech in ordinary home environments. Still, the comparative results indicate that Whisper ASR solutions perform better on kids, and we can benefit from increasing the model size from base to small, while small to medium is close.

For SLU pipeline evaluation, we test our highest-performing NLU models on noisy ASR output. Table~\ref{nlu+asr-results} presents the Intent and Entity Classification results achieved on home deployment data where the DIET+ConveRT models run on varying ASR models output. Note that Voice Activity Detection (VAD) is an integral part of ASR that decides the presence/absence of human speech. We realize that the VAD stage is filtering out a lot of audio chunks with actual kid speech with Rockhopper and Google. Thus, our VAD-ASR nodes can ignore a lot of audio segments with reference transcripts (57.9\% for Rokchopper, 49.1\% for Google). That is less of an issue with Whisper-base/small/medium, missing 7.1\%/5.7\%/4.4\% of transcribed utterances (often due to filtering very long and repetitive trash Whisper output). When we treat these entirely missed utterances with no ASR output as classification errors for NLU tasks (i.e., missing to predict intent/entities when no speech is detected), we can adjust the F1-scores accordingly to evaluate the VAD-ASR+NLU pipeline. These VAD-adjusted F1-scores are compared in Table~\ref{nlu+asr-results}, aligned with the WER results, where NLU on Whisper ASR performs relatively higher than Google and Rockhopper. For enhanced Intent Recognition in real-world deployments with kids, increasing the ASR model size from small to medium could be worth the trouble for Whisper. Yet, the F1 drop is still huge, from 94.2\% with NLU to 73.1\% with VAD-ASR+NLU, when VAD-ASR errors propagate into the SLU pipeline. 

\begin{table}[!t]
  \centering
  \small
  \resizebox{\columnwidth}{!}{
  \begin{tabular}{lcccc}
    \toprule
     & \multicolumn{2}{c}{\textbf{Intent Detection}} & \multicolumn{2}{c}{\textbf{Entity Extraction}}\\
    \textbf{ASR Model} & F1 & Adjusted-F1 & F1 & Adjusted-F1 \\
    \midrule
    Rockhopper      & 36.7 & 15.5 & 82.9 & 35.0 \\
    Google Cloud    & \textbf{78.0} & 39.7 & 96.2 & 49.0 \\
    Whisper-base    & 64.7 & 60.0 & 95.4 & 88.5 \\
    Whisper-small   & 72.2 & 68.1 & 96.6 & 91.1 \\
    Whisper-medium  & \textbf{76.5} & \textbf{73.1} & \textbf{98.5} & \textbf{94.1} \\
    \bottomrule
  \end{tabular}
  }
  \caption{SLU Pipeline Evaluation Results in F1-scores (\%) for ASR+NLU and VAD-Adjusted ASR+NLU on Kid Space Home Deployment Data}
  \label{nlu+asr-results}
\end{table}

\begin{table*}[!ht]
  \centering
  \small
  \begin{tabular}{lcc}
    \toprule
    \textbf{Sample Kid Utterance} & \textbf{Intent} & \textbf{Prediction} \\
    \midrule
    Pepper. & \textit{state-name} & \textit{answer-valid} \\
    Wow, that's a lot of red flowers. & \textit{out-of-scope} & \textit{answer-flowers} \\
    None. & \textit{state-number} & \textit{deny} \\
    Nothing. & \textit{state-number} & \textit{deny} \\
    Yeah. Can we have some carrots? & \textit{affirm} & \textit{out-of-scope} \\
    Okay, Do your magic. & \textit{affirm} & \textit{out-of-scope} \\
    Maybe tomorrow. & \textit{affirm} & \textit{out-of-scope} \\
    He's a bear. & \textit{out-of-scope} & \textit{answer-valid} \\
    I like the idea of a bear & \textit{out-of-scope} & \textit{answer-valid} \\
    Oh, 46? Okay. & \textit{still-counting} & \textit{state-number} \\
    94. Okay. & \textit{still-counting} & \textit{state-number} \\
    Now we have mountains. & \textit{out-of-scope} & \textit{answer-valid} \\
    A pond? & \textit{out-of-scope} & \textit{answer-valid} \\
    Sorry, I didn't understand it. Uh, five tens. & \textit{state-number} & \textit{still-counting} \\
    Ah this is 70, 7. & \textit{state-number} & \textit{still-counting} \\
    \bottomrule
  \end{tabular}
  \caption{NLU Error Analysis: Intent Recognition Error Samples from Kid Space Home Deployment Data}
  \label{nlu-errors}
\end{table*}

\begin{table*}[!ht]
  \centering
  \small
  \begin{tabular}{lllcc}
    \toprule
    \textbf{Human Transcript} & \textbf{ASR Output} & \textbf{ASR Model} & \textbf{Intent} & \textbf{Prediction} \\
    \midrule
    Six. & thanks & Rockhopper & \textit{state-number} & \textit{thank} \\
    fifteen & if he & Rockhopper & \textit{state-number} & \textit{out-of-scope} \\
    \midrule
    fifteen & Mickey & Google Cloud & \textit{state-number} & \textit{state-name} \\
    Five. & bye & Google Cloud & \textit{state-number} & \textit{goodbye} \\
    \midrule
    Blue. & Blair. & Whisper-base & \textit{state-color} & \textit{state-name} \\
    twenty & Plenty. & Whisper-base & \textit{state-number} & \textit{had-fun-a-lot} \\
    A lot. & Oh, la. & Whisper-base & \textit{had-fun-a-lot} & \textit{out-of-scope} \\
    \midrule
    A lot. & Oh, wow. & Whisper-small & \textit{had-fun-a-lot} & \textit{out-of-scope} \\
    Two. & you & Whisper-small & \textit{state-number} & \textit{out-of-scope} \\
    Four. & I'm going to see this floor. & Whisper-small & \textit{state-number} & \textit{out-of-scope} \\
    \midrule
    twenty & Swamy? & Whisper-medium & \textit{state-number} & \textit{state-name} \\
    Eight. & E. & Whisper-medium & \textit{state-number} & \textit{out-of-scope} \\
    \bottomrule
  \end{tabular}
  \caption{SLU Pipeline (ASR+NLU): Intent Recognition Error Samples from Kid Space Home Deployment Data}
  \label{asr+nlu-errors}
\end{table*}

\section{Error Analysis}

For NLU error analysis, Table~\ref{nlu-errors} reveals utterance samples from our Kid Space Home Deployment data with misclassified intents obtained by the DIET+ConveRT models on manual/human transcripts. These language understanding errors illustrate the potential pain points solely related to the NLU model performances, as we are assuming perfect or human-level ASR here by feeding the manually transcribed utterances into the NLU. Such intent prediction errors occur in real-world deployments for many reasons. For example, authentic user utterances can have multiple intents (e.g., ``\textit{Yeah. Can we have some carrots?}'' starts with \textit{affirm} and continues with \textit{out-of-scope}). Some utterances can be challenging due to subtle differences between intent classes (e.g., ``\textit{Ah this is 70, 7.}'' is submitting a verbal answer with \textit{state-number} but can easily be mixed with \textit{still-counting} too). Moreover, we observe utterances having \textit{colors} and ``\textit{flowers}'' within \textit{out-of-scope} (e.g., ``\textit{Wow, that’s a lot of red flowers.}''), which can be confusing for the NLU models trained on cleaner POC datasets. 

For further error analysis on the SLU pipeline (ASR+NLU), Table~\ref{asr+nlu-errors} demonstrates Intent Recognition error samples from Kid Space Home Deployment data obtained on ASR output with several speech recognition models we explored. These samples depict anticipated error propagation from speech recognition to language understanding modules in the cascaded SLU approach. Please check Appendix~\ref{sec:appendix} for a more detailed ASR error analysis.

\section{Conclusion}

To increase the quality of math learning experiences at home for early childhood education, we develop a multimodal dialogue system with play-based learning activities, helping the kids gain basic math skills. This study investigates a modular SLU pipeline for kids with cascading ASR and NLU modules, evaluated on our first home deployment data with 12 kids at individual homes. For NLU, we examine the advantages of a multi-task architecture and experiment with numerous pretrained language representations for Intent Recognition and Entity Extraction tasks in our application domain. For ASR, we inspect the WER with several solutions that are either low-power and local (e.g., Rockhopper), commercial (e.g., Google Cloud), or open-source (e.g., Whisper) with varying model sizes and conclude that Whisper-medium outperforms the rest on kids' speech at authentic home environments. Finally, we evaluate the SLU pipeline by running our best-performing NLU models, DIET+ConveRT, on VAD-ASR output to observe the significant effects of cascaded errors due to noisy voice detection and speech recognition performance with kids in realistic home deployment settings. In the future, we aim to fine-tune the Whisper ASR acoustic models on kids' speech and language models on domain-specific math content. Moreover, we consider exploring N-Best-ASR-Transformers~\cite{ganesan-etal-2021-n} to leverage multiple Whisper ASR hypotheses and mitigate errors propagated into cascading SLU.

\section*{Limitations}

By building this task-specific dialogue system for kids, we aim to increase the overall quality of basic math education and learning at-home experiences for younger children. In our previous school deployments, the overall cost of the whole school/classroom setup, including the wall/ceiling-mounted projector, 3D/RGB-D cameras, LiDAR sensor, wireless lavalier microphones, servers, etc., can be considered as a limitation for public schools and disadvantaged populations. When we shifted our focus to home learning usages after the COVID-19 pandemic, we simplified the overall setup for 1:1 learning with a PC laptop with a built-in camera, a depth camera on a tripod, a lapel mic, and a playmat with cubes and sticks. However, even this minimal instrumentation suitable for home setup can be a limitation for kids with lower socioeconomic status. Moreover, the dataset size of our initial home deployment data collected from 12 kids in 12 sessions is relatively small, with around 12 hours of audio data manually transcribed and annotated. Collecting multimodal data at authentic homes of individual kids within our target age group (e.g., 5-to-8 years old) and labor-intensive labeling process is challenging and costly. To overcome these data scarcity limitations and develop dialogue systems for kids with such small-data regimes, we had to rely on transfer learning approaches as much as possible. However, the dataset sizes affect the generalizability of our explorations, the reliability of some results, and ultimately the robustness of our multimodal dialogue system for deployments with kids in the real world.


\section*{Ethics Statement}

Prior to our initial research deployments at home, a meticulous process of Privacy Impact Assessment is pursued. The legal approval processes are completed to operate our research with educators, parents, and the kids. Individual participants and parties involved have signed the relevant consent forms in advance, which inform essential details about our research studies. The intentions and procedures and how the participant data will be collected and utilized to facilitate our research are explained in writing in these required consent forms. Our collaborators comply with stricter data privacy policies as well.


\section*{Acknowledgements}


We aspire to share our gratitude and acknowledge our former and current colleagues in the Kid Space team at Intel Labs. Particularly: (i) Hector Cordourier Maruri, Juan Del Hoyo Ontiveros, and Georg Stemmer for developing the VAD-ASR node to obtain the ASR output that we use in our SLU pipeline; (ii) Benjamin Bair, Lenitra Durham, Sai Prasad, Giuseppe Raffa, Celal Savur, and Sangita Sharma for designing the HW/SW architectural setup, developing the Wizard UI and game logic nodes, and supporting data collection; (iii) Ankur Agrawal, Arturo Bringas Garcia, Vishwajeet Narwal, and Guillermo Rivas Aguilar for developing the Student UI via the Unity game engine; (iv) Glen Anderson, Sinem Aslan, Rebecca Chierichetti, Pete Denman, John Sherry, and Meng Shi for conducting UX studies and performing interaction design to conceptualize school and home usages; (v) David Gonzalez Aguirre, Gesem Gudino Mejia, and Julio Zamora Esquivel for developing the visual understanding nodes to support this research. We would also want to gratefully acknowledge our field team members from Summa Linguae Technologies, especially Rick Lin and Brenda Tumbalobos Cubas, for their exceptional support in executing the data collection and transcription/annotation tasks in collaboration with our Intel Labs Kid Space team. Finally, we should thank the Rasa team and community developers for their open-source framework and contributions that empowered us to conduct our research.

\bibliography{anthology,custom}

\begin{thebibliography}{113}
\expandafter\ifx\csname natexlab\endcsname\relax\def\natexlab#1{#1}\fi

\bibitem[{Anderson et~al.(2018)Anderson, Panneer, Shi, Marshall, Agrawal,
  Chierichetti, Raffa, Sherry, Loi, and Durham}]{KidSpace-ICMI-2018}
Glen~J. Anderson, Selvakumar Panneer, Meng Shi, Carl~S. Marshall, Ankur
  Agrawal, Rebecca Chierichetti, Giuseppe Raffa, John Sherry, Daria Loi, and
  Lenitra~Megail Durham. 2018.
\newblock \href {https://doi.org/10.1145/3279981.3279986} {Kid space:
  Interactive learning in a smart environment}.
\newblock In \emph{Proceedings of the Group Interaction Frontiers in
  Technology}, GIFT'18, New York, NY, USA. Association for Computing Machinery.

\bibitem[{Aslan et~al.(2022)Aslan, Agrawal, Alyuz, Chierichetti, Durham,
  Manuvinakurike, Okur, Sahay, Sharma, Sherry, Raffa, and
  Nachman}]{KidSpace-ETRD-2022}
Sinem Aslan, Ankur Agrawal, Nese Alyuz, Rebecca Chierichetti, Lenitra~M Durham,
  Ramesh Manuvinakurike, Eda Okur, Saurav Sahay, Sangita Sharma, John Sherry,
  Giuseppe Raffa, and Lama Nachman. 2022.
\newblock \href {https://doi.org/10.1007/s11423-021-10072-x} {Exploring kid
  space in the wild: a preliminary study of multimodal and immersive
  collaborative play-based learning experiences}.
\newblock \emph{Educational Technology Research and Development}, 70:205--230.

\bibitem[{Aslan et~al.(2019)Aslan, Alyuz, Tanriover, Mete, Okur, D'Mello, and
  Arslan~Esme}]{AdaptLE-CHI-2019}
Sinem Aslan, Nese Alyuz, Cagri Tanriover, Sinem~E. Mete, Eda Okur, Sidney~K.
  D'Mello, and Asli Arslan~Esme. 2019.
\newblock \href {https://doi.org/10.1145/3290605.3300534} {Investigating the
  impact of a real-time, multimodal student engagement analytics technology in
  authentic classrooms}.
\newblock In \emph{Proceedings of the 2019 CHI Conference on Human Factors in
  Computing Systems}, CHI '19, page 1–12, New York, NY, USA. Association for
  Computing Machinery.

\bibitem[{Azerbayev et~al.(2022)Azerbayev, Piotrowski, and
  Avigad}]{azerbayev2022proofnet}
Zhangir Azerbayev, Bartosz Piotrowski, and Jeremy Avigad. 2022.
\newblock \href {https://mathai2022.github.io/papers/20.pdf} {Proofnet: A
  benchmark for autoformalizing and formally proving undergraduate-level
  mathematics problems}.
\newblock In \emph{Workshop MATH-AI: Toward Human-Level Mathematical Reasoning,
  36th Conference on Neural Information Processing Systems (NeurIPS 2022), New
  Orleans, Louisiana, USA}.

\bibitem[{Baevski et~al.(2020)Baevski, Zhou, Mohamed, and
  Auli}]{baevski2020wav2vec}
Alexei Baevski, Yuhao Zhou, Abdelrahman Mohamed, and Michael Auli. 2020.
\newblock wav2vec 2.0: A framework for self-supervised learning of speech
  representations.
\newblock \emph{Advances in neural information processing systems},
  33:12449--12460.

\bibitem[{Bai et~al.(2021)Bai, Hubers, Cucchiarini, and Strik}]{Bai2021}
Yu~Bai, Ferdy Hubers, Catia Cucchiarini, and Helmer Strik. 2021.
\newblock \href {https://doi.org/10.21437/IberSPEECH.2021-3} {{An ASR-based
  Reading Tutor for Practicing Reading Skills in the First Grade: Improving
  Performance through Threshold Adjustment}}.
\newblock In \emph{Proc. IberSPEECH 2021}, pages 11--15.

\bibitem[{Bai et~al.(2022)Bai, Hubers, Cucchiarini, {van Hout}, and
  Strik}]{bai22b_interspeech}
Yu~Bai, Ferdy Hubers, Catia Cucchiarini, Roeland {van Hout}, and Helmer Strik.
  2022.
\newblock \href {https://doi.org/10.21437/Interspeech.2022-10810} {{The Effects
  of Implicit and Explicit Feedback in an ASR-based Reading Tutor for Dutch
  First-graders}}.
\newblock In \emph{Proc. Interspeech 2022}, pages 4476--4480.

\bibitem[{Baker(2021)}]{baker2021artificial}
Ryan~S Baker. 2021.
\newblock Artificial intelligence in education: Bringing it all together.
\newblock \emph{OECD Digital Education Outlook 2021: Pushing the Frontiers with
  Artificial Intelligence, Blockchain and Robots}, pages 43--51.

\bibitem[{Bhardwaj et~al.(2022)Bhardwaj, Ben~Othman, Kukreja, Belkhier, Bajaj,
  Goud, Rehman, Shafiq, and Hamam}]{bhardwaj2022automatic}
Vivek Bhardwaj, Mohamed~Tahar Ben~Othman, Vinay Kukreja, Youcef Belkhier, Mohit
  Bajaj, B~Srikanth Goud, Ateeq~Ur Rehman, Muhammad Shafiq, and Habib Hamam.
  2022.
\newblock Automatic speech recognition (asr) systems for children: A systematic
  literature review.
\newblock \emph{Applied Sciences}, 12(9):4419.

\bibitem[{Bibauw et~al.(2022)Bibauw, Van~den Noortgate, Fran{\c{c}}ois, and
  Desmet}]{bibauw2022dialogue}
Serge Bibauw, Wim Van~den Noortgate, Thomas Fran{\c{c}}ois, and Piet Desmet.
  2022.
\newblock Dialogue systems for language learning: a meta-analysis.
\newblock \emph{Language Learning \& Technology}, 26(1).

\bibitem[{Blanchard et~al.(2015)Blanchard, Brady, Olney, Glaus, Sun, Nystrand,
  Samei, Kelly, and D'Mello}]{10.1007/978-3-319-19773-9_3}
Nathaniel Blanchard, Michael Brady, Andrew~M. Olney, Marci Glaus, Xiaoyi Sun,
  Martin Nystrand, Borhan Samei, Sean Kelly, and Sidney D'Mello. 2015.
\newblock A study of automatic speech recognition in noisy classroom
  environments for automated dialog analysis.
\newblock In \emph{Artificial Intelligence in Education}, pages 23--33, Cham.
  Springer International Publishing.

\bibitem[{Bocklisch et~al.(2017)Bocklisch, Faulkner, Pawlowski, and
  Nichol}]{bocklisch2017rasa}
Tom Bocklisch, Joey Faulkner, Nick Pawlowski, and Alan Nichol. 2017.
\newblock \href {http://arxiv.org/abs/arXiv:1712.05181} {Rasa: Open source
  language understanding and dialogue management}.
\newblock In \emph{Conversational AI Workshop, NIPS 2017}.

\bibitem[{Booth et~al.(2020)Booth, Carns, Kennington, and
  Rafla}]{booth-etal-2020-evaluating}
Eric Booth, Jake Carns, Casey Kennington, and Nader Rafla. 2020.
\newblock \href {https://aclanthology.org/2020.lrec-1.778} {Evaluating and
  improving child-directed automatic speech recognition}.
\newblock In \emph{Proceedings of the Twelfth Language Resources and Evaluation
  Conference}, pages 6340--6345, Marseille, France. European Language Resources
  Association.

\bibitem[{Brown et~al.(2020)Brown, Mann, Ryder, Subbiah, Kaplan, Dhariwal,
  Neelakantan, Shyam, Sastry, Askell et~al.}]{brown2020language}
Tom Brown, Benjamin Mann, Nick Ryder, Melanie Subbiah, Jared~D Kaplan, Prafulla
  Dhariwal, Arvind Neelakantan, Pranav Shyam, Girish Sastry, Amanda Askell,
  et~al. 2020.
\newblock Language models are few-shot learners.
\newblock \emph{Advances in neural information processing systems},
  33:1877--1901.

\bibitem[{Budzianowski et~al.(2018)Budzianowski, Wen, Tseng, Casanueva, Ultes,
  Ramadan, and Ga{\v{s}}i{\'c}}]{budzianowski-etal-2018-multiwoz}
Pawe{\l} Budzianowski, Tsung-Hsien Wen, Bo-Hsiang Tseng, I{\~n}igo Casanueva,
  Stefan Ultes, Osman Ramadan, and Milica Ga{\v{s}}i{\'c}. 2018.
\newblock \href {https://doi.org/10.18653/v1/D18-1547} {{M}ulti{WOZ} - a
  large-scale multi-domain {W}izard-of-{O}z dataset for task-oriented dialogue
  modelling}.
\newblock In \emph{Proceedings of the 2018 Conference on Empirical Methods in
  Natural Language Processing}, pages 5016--5026, Brussels, Belgium.
  Association for Computational Linguistics.

\bibitem[{Bunk et~al.(2020)Bunk, Varshneya, Vlasov, and Nichol}]{DIET-2020}
Tanja Bunk, Daksh Varshneya, Vladimir Vlasov, and Alan Nichol. 2020.
\newblock \href {http://arxiv.org/abs/2004.09936} {{DIET:} lightweight language
  understanding for dialogue systems}.
\newblock \emph{CoRR}, abs/2004.09936.

\bibitem[{Burtsev et~al.(2018)Burtsev, Seliverstov, Airapetyan, Arkhipov,
  Baymurzina, Bushkov, Gureenkova, Khakhulin, Kuratov, Kuznetsov, Litinsky,
  Logacheva, Lymar, Malykh, Petrov, Polulyakh, Pugachev, Sorokin, Vikhreva, and
  Zaynutdinov}]{burtsev-etal-2018-deeppavlov}
Mikhail Burtsev, Alexander Seliverstov, Rafael Airapetyan, Mikhail Arkhipov,
  Dilyara Baymurzina, Nickolay Bushkov, Olga Gureenkova, Taras Khakhulin, Yuri
  Kuratov, Denis Kuznetsov, Alexey Litinsky, Varvara Logacheva, Alexey Lymar,
  Valentin Malykh, Maxim Petrov, Vadim Polulyakh, Leonid Pugachev, Alexey
  Sorokin, Maria Vikhreva, and Marat Zaynutdinov. 2018.
\newblock \href {https://doi.org/10.18653/v1/P18-4021} {{D}eep{P}avlov:
  Open-source library for dialogue systems}.
\newblock In \emph{Proceedings of {ACL} 2018, System Demonstrations}, pages
  122--127, Melbourne, Australia. Association for Computational Linguistics.

\bibitem[{Cahill et~al.(2020)Cahill, Fife, Riordan, Vajpayee, and
  Galochkin}]{cahill-etal-2020-context}
Aoife Cahill, James~H Fife, Brian Riordan, Avijit Vajpayee, and Dmytro
  Galochkin. 2020.
\newblock \href {https://doi.org/10.18653/v1/2020.bea-1.19} {Context-based
  automated scoring of complex mathematical responses}.
\newblock In \emph{Proceedings of the Fifteenth Workshop on Innovative Use of
  NLP for Building Educational Applications}, pages 186--192, Seattle, WA, USA
  → Online. Association for Computational Linguistics.

\bibitem[{Cesarone(2008)}]{cesarone2008early}
Bernard Cesarone. 2008.
\newblock Early childhood mathematics: Promoting good beginnings.
\newblock \emph{Childhood Education}, 84(3):189.

\bibitem[{Chan et~al.(2021)Chan, Chung, and
  Fan}]{DBLP:journals/corr/abs-2112-01012}
Ying{-}Hong Chan, Ho{-}Lam Chung, and Yao{-}Chung Fan. 2021.
\newblock \href {http://arxiv.org/abs/2112.01012} {Improving controllability of
  educational question generation by keyword provision}.
\newblock \emph{CoRR}, abs/2112.01012.

\bibitem[{Chassignol et~al.(2018)Chassignol, Khoroshavin, Klimova, and
  Bilyatdinova}]{CHASSIGNOL201816}
Maud Chassignol, Aleksandr Khoroshavin, Alexandra Klimova, and Anna
  Bilyatdinova. 2018.
\newblock \href {https://doi.org/https://doi.org/10.1016/j.procs.2018.08.233}
  {Artificial intelligence trends in education: a narrative overview}.
\newblock \emph{Procedia Computer Science}, 136:16--24.
\newblock 7th International Young Scientists Conference on Computational
  Science, YSC2018, 02-06 July2018, Heraklion, Greece.

\bibitem[{Claus et~al.(2013)Claus, Rosales, Petrick, Hain, and
  Hoffmann}]{claus2013survey}
Felix Claus, Hamurabi~Gamboa Rosales, Rico Petrick, Horst-Udo Hain, and
  R{\"u}diger Hoffmann. 2013.
\newblock A survey about databases of children's speech.
\newblock In \emph{INTERSPEECH}, pages 2410--2414.

\bibitem[{Cotton et~al.(2023)Cotton, Cotton, and Shipway}]{cotton2023chatting}
Debby~RE Cotton, Peter~A Cotton, and J~Reuben Shipway. 2023.
\newblock Chatting and cheating: Ensuring academic integrity in the era of
  chatgpt.
\newblock \emph{Innovations in Education and Teaching International}, pages
  1--12.

\bibitem[{Dai et~al.(2019)Dai, Yang, Yang, Carbonell, Le, and
  Salakhutdinov}]{dai-etal-2019-transformer}
Zihang Dai, Zhilin Yang, Yiming Yang, Jaime Carbonell, Quoc Le, and Ruslan
  Salakhutdinov. 2019.
\newblock \href {https://doi.org/10.18653/v1/P19-1285} {Transformer-{XL}:
  Attentive language models beyond a fixed-length context}.
\newblock In \emph{Proceedings of the 57th Annual Meeting of the Association
  for Computational Linguistics}, pages 2978--2988, Florence, Italy.
  Association for Computational Linguistics.

\bibitem[{Datta et~al.(2020)Datta, Phillips, Chiu, Watson, Bywater, Barnes, and
  Brown}]{DBLP:journals/corr/abs-2010-12710}
Debajyoti Datta, Maria Phillips, Jennifer~L. Chiu, Ginger~S. Watson, James~P.
  Bywater, Laura~E. Barnes, and Donald~E. Brown. 2020.
\newblock \href {http://arxiv.org/abs/2010.12710} {Improving classification
  through weak supervision in context-specific conversational agent development
  for teacher education}.
\newblock \emph{CoRR}, abs/2010.12710.

\bibitem[{Devlin et~al.(2019)Devlin, Chang, Lee, and
  Toutanova}]{devlin-etal-2019-bert}
Jacob Devlin, Ming-Wei Chang, Kenton Lee, and Kristina Toutanova. 2019.
\newblock \href {https://doi.org/10.18653/v1/N19-1423} {{BERT}: Pre-training of
  deep bidirectional transformers for language understanding}.
\newblock In \emph{Proceedings of the 2019 Conference of the North {A}merican
  Chapter of the Association for Computational Linguistics: Human Language
  Technologies, Volume 1 (Long and Short Papers)}, pages 4171--4186,
  Minneapolis, Minnesota. Association for Computational Linguistics.

\bibitem[{Duan and Chen(2020)}]{duan2020unsupervised}
Richeng Duan and Nancy~F Chen. 2020.
\newblock Unsupervised feature adaptation using adversarial multi-task training
  for automatic evaluation of children's speech.
\newblock In \emph{INTERSPEECH}, pages 3037--3041.

\bibitem[{Dutta et~al.(2022)Dutta, Irvin, Buzhardt, and
  Hansen}]{dutta-etal-2022-activity}
Satwik Dutta, Dwight Irvin, Jay Buzhardt, and John~H.L. Hansen. 2022.
\newblock \href {https://doi.org/10.18653/v1/2022.bea-1.13} {Activity focused
  speech recognition of preschool children in early childhood classrooms}.
\newblock In \emph{Proceedings of the 17th Workshop on Innovative Use of NLP
  for Building Educational Applications (BEA 2022)}, pages 92--100, Seattle,
  Washington. Association for Computational Linguistics.

\bibitem[{Falmagne et~al.(2013)Falmagne, Albert, Doble, Eppstein, and
  Hu}]{falmagne2013knowledge}
Jean-Claude Falmagne, Dietrich Albert, Christopher Doble, David Eppstein, and
  Xiangen Hu. 2013.
\newblock \emph{Knowledge spaces: Applications in education}.
\newblock Springer Science \& Business Media.

\bibitem[{Feng et~al.(2022)Feng, Yang, Cer, Arivazhagan, and
  Wang}]{feng-etal-2022-language}
Fangxiaoyu Feng, Yinfei Yang, Daniel Cer, Naveen Arivazhagan, and Wei Wang.
  2022.
\newblock \href {https://doi.org/10.18653/v1/2022.acl-long.62}
  {Language-agnostic {BERT} sentence embedding}.
\newblock In \emph{Proceedings of the 60th Annual Meeting of the Association
  for Computational Linguistics (Volume 1: Long Papers)}, pages 878--891,
  Dublin, Ireland. Association for Computational Linguistics.

\bibitem[{Franck~Dernoncourt(2018)}]{ASRbenchmark2018}
Walter~Chang Franck~Dernoncourt, Trung~Bui. 2018.
\newblock A framework for speech recognition benchmarking.
\newblock In \emph{Interspeech}.

\bibitem[{Frieder et~al.(2023)Frieder, Pinchetti, Griffiths, Salvatori,
  Lukasiewicz, Petersen, Chevalier, and Berner}]{frieder2023mathematical}
Simon Frieder, Luca Pinchetti, Ryan-Rhys Griffiths, Tommaso Salvatori, Thomas
  Lukasiewicz, Philipp~Christian Petersen, Alexis Chevalier, and Julius Berner.
  2023.
\newblock Mathematical capabilities of chatgpt.
\newblock \emph{arXiv preprint arXiv:2301.13867}.

\bibitem[{Ganesan et~al.(2021)Ganesan, Bamdev, B, Venugopal, and
  Tushar}]{ganesan-etal-2021-n}
Karthik Ganesan, Pakhi Bamdev, Jaivarsan B, Amresh Venugopal, and Abhinav
  Tushar. 2021.
\newblock \href {https://doi.org/10.18653/v1/2021.acl-short.14} {N-best {ASR}
  transformer: Enhancing {SLU} performance using multiple {ASR} hypotheses}.
\newblock In \emph{Proceedings of the 59th Annual Meeting of the Association
  for Computational Linguistics and the 11th International Joint Conference on
  Natural Language Processing (Volume 2: Short Papers)}, pages 93--98, Online.
  Association for Computational Linguistics.

\bibitem[{Gerosa et~al.(2007)Gerosa, Giuliani, and
  Brugnara}]{gerosa2007acoustic}
Matteo Gerosa, Diego Giuliani, and Fabio Brugnara. 2007.
\newblock Acoustic variability and automatic recognition of children’s
  speech.
\newblock \emph{Speech Communication}, 49(10-11):847--860.

\bibitem[{Goo et~al.(2018)Goo, Gao, Hsu, Huo, Chen, Hsu, and
  Chen}]{goo-etal-2018-slot}
Chih-Wen Goo, Guang Gao, Yun-Kai Hsu, Chih-Li Huo, Tsung-Chieh Chen, Keng-Wei
  Hsu, and Yun-Nung Chen. 2018.
\newblock \href {https://doi.org/10.18653/v1/N18-2118} {Slot-gated modeling for
  joint slot filling and intent prediction}.
\newblock In \emph{Proceedings of the 2018 Conference of the North {A}merican
  Chapter of the Association for Computational Linguistics: Human Language
  Technologies, Volume 2 (Short Papers)}, pages 753--757, New Orleans,
  Louisiana. Association for Computational Linguistics.

\bibitem[{Graesser et~al.(2005)Graesser, Chipman, Haynes, and Olney}]{1532370}
A.C. Graesser, P.~Chipman, B.C. Haynes, and A.~Olney. 2005.
\newblock \href {https://doi.org/10.1109/TE.2005.856149} {Autotutor: an
  intelligent tutoring system with mixed-initiative dialogue}.
\newblock \emph{IEEE Transactions on Education}, 48(4):612--618.

\bibitem[{Grossman et~al.(2019)Grossman, Lin, Sheng, Wei, Williams, and
  Goel}]{grossman2019mathbot}
Joshua Grossman, Zhiyuan Lin, Hao Sheng, Johnny Tian-Zheng Wei, Joseph~J
  Williams, and Sharad Goel. 2019.
\newblock Mathbot: Transforming online resources for learning math into
  conversational interactions.
\newblock \emph{AAAI 2019 Story-Enabled Intelligence}.

\bibitem[{Henderson et~al.(2020)Henderson, Casanueva, Mrk{\v{s}}i{\'c}, Su,
  Wen, and Vuli{\'c}}]{ConveRT-2020}
Matthew Henderson, I{\~n}igo Casanueva, Nikola Mrk{\v{s}}i{\'c}, Pei-Hao Su,
  Tsung-Hsien Wen, and Ivan Vuli{\'c}. 2020.
\newblock \href {https://doi.org/10.18653/v1/2020.findings-emnlp.196}
  {{C}onve{RT}: Efficient and accurate conversational representations from
  transformers}.
\newblock In \emph{Findings of the Association for Computational Linguistics:
  EMNLP 2020}, pages 2161--2174, Online. Association for Computational
  Linguistics.

\bibitem[{Honnibal et~al.(2020)Honnibal, Montani, Van~Landeghem, and
  Boyd}]{honnibal10boyd}
Matthew Honnibal, Ines Montani, Sofie Van~Landeghem, and Adriane Boyd. 2020.
\newblock \href {https://doi.org/10.5281/zenodo.1212303} {{spaCy}:
  Industrial-strength natural language processing in python}.

\bibitem[{Huang et~al.(2021)Huang, Wang, Xu, Cao, and Yang}]{huangreal2}
Shifeng Huang, Jiawei Wang, Jiao Xu, Da~Cao, and Ming Yang. 2021.
\newblock \href {https://mathai4ed.github.io/papers/papers/paper_7.pdf} {Real2:
  An end-to-end memory-augmented solver for math word problems}.
\newblock In \emph{Workshop on Math AI for Education (MATHAI4ED), 35th
  Conference on Neural Information Processing Systems (NeurIPS 2021)}.

\bibitem[{Jia et~al.(2020)Jia, He, and Le}]{MMHCI}
Jiyou Jia, Yunfan He, and Huixiao Le. 2020.
\newblock A multimodal human-computer interaction system and its application in
  smart learning environments.
\newblock In \emph{Blended Learning. Education in a Smart Learning
  Environment}, pages 3--14, Cham. Springer International Publishing.

\bibitem[{Kasneci et~al.(2023)Kasneci, Se{\ss}ler, K{\"u}chemann, Bannert,
  Dementieva, Fischer, Gasser, Groh, G{\"u}nnemann, H{\"u}llermeier
  et~al.}]{kasneci2023chatgpt}
Enkelejda Kasneci, Kathrin Se{\ss}ler, Stefan K{\"u}chemann, Maria Bannert,
  Daryna Dementieva, Frank Fischer, Urs Gasser, Georg Groh, Stephan
  G{\"u}nnemann, Eyke H{\"u}llermeier, et~al. 2023.
\newblock Chatgpt for good? on opportunities and challenges of large language
  models for education.
\newblock \emph{Learning and Individual Differences}, 103:102274.

\bibitem[{Kelly et~al.(2020)Kelly, Karamichali, Saeb, Vesel{\`y}, Parslow,
  Deng, Letondor, O'Regan, and Zhou}]{kelly2020soapbox}
Amelia~C Kelly, Eleni Karamichali, Armin Saeb, Karel Vesel{\`y}, Nicholas
  Parslow, Agape Deng, Arnaud Letondor, Robert O'Regan, and Qiru Zhou. 2020.
\newblock Soapbox labs verification platform for child speech.
\newblock In \emph{INTERSPEECH}, pages 486--487.

\bibitem[{Lafferty et~al.(2001)Lafferty, McCallum, and
  Pereira}]{DBLP:conf/icml/LaffertyMP01}
John~D. Lafferty, Andrew McCallum, and Fernando C.~N. Pereira. 2001.
\newblock Conditional random fields: Probabilistic models for segmenting and
  labeling sequence data.
\newblock In \emph{International Conference on Machine Learning}, ICML, pages
  282--289.

\bibitem[{Latif et~al.(2023)Latif, Zaidi, Cuayahuitl, Shamshad, Shoukat, and
  Qadir}]{latif2023transformers}
Siddique Latif, Aun Zaidi, Heriberto Cuayahuitl, Fahad Shamshad, Moazzam
  Shoukat, and Junaid Qadir. 2023.
\newblock Transformers in speech processing: A survey.
\newblock \emph{arXiv preprint arXiv:2303.11607}.

\bibitem[{Lende and Raghuwanshi(2016)}]{lende2016question}
Sweta~P Lende and MM~Raghuwanshi. 2016.
\newblock Question answering system on education acts using nlp techniques.
\newblock In \emph{2016 world conference on futuristic trends in research and
  innovation for social welfare (Startup Conclave)}, pages 1--6. IEEE.

\bibitem[{Liu and Lane(2016)}]{Liu+2016}
Bing Liu and Ian Lane. 2016.
\newblock \href {https://doi.org/10.21437/Interspeech.2016-1352}
  {Attention-based recurrent neural network models for joint intent detection
  and slot filling}.
\newblock In \emph{Interspeech 2016}, pages 685--689.

\bibitem[{Liu et~al.(2021{\natexlab{a}})Liu, Zheng, Du, Ding, Qian, Yang, and
  Tang}]{liu2021gpt}
Xiao Liu, Yanan Zheng, Zhengxiao Du, Ming Ding, Yujie Qian, Zhilin Yang, and
  Jie Tang. 2021{\natexlab{a}}.
\newblock Gpt understands, too.
\newblock \emph{arXiv preprint arXiv:2103.10385}.

\bibitem[{Liu et~al.(2021{\natexlab{b}})Liu, Eshghi, Swietojanski, and
  Rieser}]{liu2021benchmarking}
Xingkun Liu, Arash Eshghi, Pawel Swietojanski, and Verena Rieser.
  2021{\natexlab{b}}.
\newblock Benchmarking natural language understanding services for building
  conversational agents.
\newblock In \emph{Increasing Naturalness and Flexibility in Spoken Dialogue
  Interaction: 10th International Workshop on Spoken Dialogue Systems}, pages
  165--183. Springer.

\bibitem[{Liu et~al.(2019)Liu, Ott, Goyal, Du, Joshi, Chen, Levy, Lewis,
  Zettlemoyer, and Stoyanov}]{liu2019roberta}
Yinhan Liu, Myle Ott, Naman Goyal, Jingfei Du, Mandar Joshi, Danqi Chen, Omer
  Levy, Mike Lewis, Luke Zettlemoyer, and Veselin Stoyanov. 2019.
\newblock \href {http://arxiv.org/abs/1907.11692} {Roberta: {A} robustly
  optimized {BERT} pretraining approach}.
\newblock \emph{CoRR}, abs/1907.11692.

\bibitem[{Loginova and Benoit(2022)}]{loginova2022structural}
Ekaterina Loginova and Dries Benoit. 2022.
\newblock Structural information in mathematical formulas for exercise
  difficulty prediction: a comparison of nlp representations.
\newblock In \emph{Proceedings of the 17th Workshop on Innovative Use of NLP
  for Building Educational Applications (BEA 2022)}, pages 101--106.

\bibitem[{Macina et~al.(2023)Macina, Daheim, Wang, Sinha, Kapur, Gurevych, and
  Sachan}]{macina2023opportunities}
Jakub Macina, Nico Daheim, Lingzhi Wang, Tanmay Sinha, Manu Kapur, Iryna
  Gurevych, and Mrinmaya Sachan. 2023.
\newblock Opportunities and challenges in neural dialog tutoring.
\newblock \emph{arXiv preprint arXiv:2301.09919}.

\bibitem[{Madotto et~al.(2020)Madotto, Liu, Lin, and
  Fung}]{madotto2020language}
Andrea Madotto, Zihan Liu, Zhaojiang Lin, and Pascale Fung. 2020.
\newblock Language models as few-shot learner for task-oriented dialogue
  systems.
\newblock \emph{arXiv preprint arXiv:2008.06239}.

\bibitem[{Mansouri et~al.(2022)Mansouri, Novotn{\'y}, Agarwal, Oard, and
  Zanibbi}]{10.1007/978-3-031-13643-6_20}
Behrooz Mansouri, V{\'i}t Novotn{\'y}, Anurag Agarwal, Douglas~W. Oard, and
  Richard Zanibbi. 2022.
\newblock Overview of arqmath-3 (2022): Third clef lab on answer retrieval
  for questions on math.
\newblock In \emph{Experimental IR Meets Multilinguality, Multimodality, and
  Interaction}, pages 286--310, Cham. Springer International Publishing.

\bibitem[{Mansouri et~al.(2019)Mansouri, Rohatgi, Oard, Wu, Giles, and
  Zanibbi}]{mansouri2019tangent}
Behrooz Mansouri, Shaurya Rohatgi, Douglas~W Oard, Jian Wu, C~Lee Giles, and
  Richard Zanibbi. 2019.
\newblock Tangent-cft: An embedding model for mathematical formulas.
\newblock In \emph{Proceedings of the 2019 ACM SIGIR international conference
  on theory of information retrieval}, pages 11--18.

\bibitem[{Mehri et~al.(2020)Mehri, Eric, and
  Hakkani{-}T{\"{u}}r}]{DBLP:journals/corr/abs-2009-13570}
Shikib Mehri, Mihail Eric, and Dilek Hakkani{-}T{\"{u}}r. 2020.
\newblock \href {http://arxiv.org/abs/2009.13570} {Dialoglue: {A} natural
  language understanding benchmark for task-oriented dialogue}.
\newblock \emph{CoRR}, abs/2009.13570.

\bibitem[{Nrupatunga et~al.(2021)Nrupatunga, Kumar, and
  Rajagopal}]{kumarphygital}
Nrupatunga, Aashish Kumar, and Anoop Rajagopal. 2021.
\newblock \href {https://mathai4ed.github.io/papers/papers/paper_5.pdf}
  {Phygital math learning with handwriting for kids}.
\newblock In \emph{Workshop on Math AI for Education (MATHAI4ED), 35th
  Conference on Neural Information Processing Systems (NeurIPS 2021)}.

\bibitem[{Nye et~al.(2018)Nye, Pavlik, Windsor, Olney, Hajeer, and
  Hu}]{nye2018skope}
Benjamin~D Nye, Philip~I Pavlik, Alistair Windsor, Andrew~M Olney, Mustafa
  Hajeer, and Xiangen Hu. 2018.
\newblock Skope-it (shareable knowledge objects as portable intelligent
  tutors): overlaying natural language tutoring on an adaptive learning system
  for mathematics.
\newblock \emph{International journal of STEM education}, 5:1--20.

\bibitem[{Okonkwo and Ade-Ibijola(2021)}]{okonkwo2021chatbots}
Chinedu~Wilfred Okonkwo and Abejide Ade-Ibijola. 2021.
\newblock Chatbots applications in education: A systematic review.
\newblock \emph{Computers and Education: Artificial Intelligence}, 2:100033.

\bibitem[{Okur et~al.(2019)Okur, Kumar, Sahay, Arslan~Esme, and
  Nachman}]{AMIE-CICLing-2019}
Eda Okur, Shachi~H. Kumar, Saurav Sahay, Asli Arslan~Esme, and Lama Nachman.
  2019.
\newblock \href {https://doi.org/10.1007/978-3-031-24340-0_25} {Natural
  language interactions in autonomous vehicles: Intent detection and slot
  filling from passenger utterances}.
\newblock In \emph{Computational Linguistics and Intelligent Text Processing},
  pages 334--350, Cham. Springer Nature Switzerland.

\bibitem[{Okur et~al.(2022{\natexlab{a}})Okur, Sahay, Fuentes~Alba, and
  Nachman}]{KidSpace-MathNLP-EMNLP-2022}
Eda Okur, Saurav Sahay, Roddy Fuentes~Alba, and Lama Nachman.
  2022{\natexlab{a}}.
\newblock \href {https://aclanthology.org/2022.mathnlp-1.7} {End-to-end
  evaluation of a spoken dialogue system for learning basic mathematics}.
\newblock In \emph{Proceedings of the 1st Workshop on Mathematical Natural
  Language Processing (MathNLP)}, pages 51--64, Abu Dhabi, United Arab Emirates
  (Hybrid). Association for Computational Linguistics.

\bibitem[{Okur et~al.(2022{\natexlab{b}})Okur, Sahay, and
  Nachman}]{KidSpace-LREC-2022}
Eda Okur, Saurav Sahay, and Lama Nachman. 2022{\natexlab{b}}.
\newblock \href {https://aclanthology.org/2022.lrec-1.437} {Data augmentation
  with paraphrase generation and entity extraction for multimodal dialogue
  system}.
\newblock In \emph{Proceedings of the Thirteenth Language Resources and
  Evaluation Conference}, pages 4114--4125, Marseille, France. European
  Language Resources Association.

\bibitem[{Okur et~al.(2022{\natexlab{c}})Okur, Sahay, and
  Nachman}]{KidSpace-GAMNLP-LREC-2022}
Eda Okur, Saurav Sahay, and Lama Nachman. 2022{\natexlab{c}}.
\newblock \href {https://aclanthology.org/2022.games-1.4} {{NLU} for game-based
  learning in real: Initial evaluations}.
\newblock In \emph{Proceedings of the 9th Workshop on Games and Natural
  Language Processing within the 13th Language Resources and Evaluation
  Conference}, pages 28--39, Marseille, France. European Language Resources
  Association.

\bibitem[{OpenAI(2022)}]{team2022chatgpt}
OpenAI. 2022.
\newblock Chatgpt: Optimizing language models for dialogue.

\bibitem[{Panayotov et~al.(2015)Panayotov, Chen, Povey, and
  Khudanpur}]{panayotov2015librispeech}
Vassil Panayotov, Guoguo Chen, Daniel Povey, and Sanjeev Khudanpur. 2015.
\newblock Librispeech: an asr corpus based on public domain audio books.
\newblock In \emph{2015 IEEE international conference on acoustics, speech and
  signal processing (ICASSP)}, pages 5206--5210. IEEE.

\bibitem[{Peng et~al.(2021)Peng, Yuan, Gao, and
  Tang}]{DBLP:journals/corr/abs-2105-00377}
Shuai Peng, Ke~Yuan, Liangcai Gao, and Zhi Tang. 2021.
\newblock \href {http://arxiv.org/abs/2105.00377} {Mathbert: {A} pre-trained
  model for mathematical formula understanding}.
\newblock \emph{CoRR}, abs/2105.00377.

\bibitem[{Pires et~al.(2019)Pires, González~Perilli, Bakała, Fleisher,
  Sansone, and Marichal}]{10.3389/feduc.2019.00081}
Ana~Cristina Pires, Fernando González~Perilli, Ewelina Bakała, Bruno
  Fleisher, Gustavo Sansone, and Sebastián Marichal. 2019.
\newblock \href {https://doi.org/10.3389/feduc.2019.00081} {Building blocks of
  mathematical learning: Virtual and tangible manipulatives lead to different
  strategies in number composition}.
\newblock \emph{Frontiers in Education}, 4.

\bibitem[{Povey et~al.(2011)Povey, Ghoshal, Boulianne, Burget, Glembek, Goel,
  Hannemann, Motlicek, Qian, Schwarz et~al.}]{povey2011kaldi}
Daniel Povey, Arnab Ghoshal, Gilles Boulianne, Lukas Burget, Ondrej Glembek,
  Nagendra Goel, Mirko Hannemann, Petr Motlicek, Yanmin Qian, Petr Schwarz,
  et~al. 2011.
\newblock The kaldi speech recognition toolkit.
\newblock In \emph{IEEE 2011 workshop on automatic speech recognition and
  understanding}, CONF. IEEE Signal Processing Society.

\bibitem[{Raamadhurai et~al.(2019)Raamadhurai, Baker, and
  Poduval}]{raamadhurai-etal-2019-curio}
Srikrishna Raamadhurai, Ryan Baker, and Vikraman Poduval. 2019.
\newblock \href {https://doi.org/10.18653/v1/W19-4435} {Curio {S}mart{C}hat : A
  system for natural language question answering for self-paced k-12 learning}.
\newblock In \emph{Proceedings of the Fourteenth Workshop on Innovative Use of
  NLP for Building Educational Applications}, pages 336--342, Florence, Italy.
  Association for Computational Linguistics.

\bibitem[{Radford et~al.(2022)Radford, Kim, Xu, Brockman, McLeavey, and
  Sutskever}]{radford2022robust}
Alec Radford, Jong~Wook Kim, Tao Xu, Greg Brockman, Christine McLeavey, and
  Ilya Sutskever. 2022.
\newblock Robust speech recognition via large-scale weak supervision.
\newblock \emph{arXiv preprint arXiv:2212.04356}.

\bibitem[{Radford et~al.(2019)Radford, Wu, Child, Luan, Amodei, Sutskever
  et~al.}]{radford2019language}
Alec Radford, Jeffrey Wu, Rewon Child, David Luan, Dario Amodei, Ilya
  Sutskever, et~al. 2019.
\newblock Language models are unsupervised multitask learners.
\newblock \emph{OpenAI blog}, 1(8):9.

\bibitem[{Rathod et~al.(2022)Rathod, Tu, and
  Stasaski}]{rathod-etal-2022-educational}
Manav Rathod, Tony Tu, and Katherine Stasaski. 2022.
\newblock \href {https://doi.org/10.18653/v1/2022.bea-1.26} {Educational
  multi-question generation for reading comprehension}.
\newblock In \emph{Proceedings of the 17th Workshop on Innovative Use of NLP
  for Building Educational Applications (BEA 2022)}, pages 216--223, Seattle,
  Washington. Association for Computational Linguistics.

\bibitem[{Reeder et~al.(2015)Reeder, Shapiro, Wakefield, and
  D'Silva}]{reeder2015speech}
Kenneth Reeder, Jon Shapiro, Jane Wakefield, and Reg D'Silva. 2015.
\newblock Speech recognition software contributes to reading development for
  young learners of english.
\newblock \emph{International Journal of Computer-Assisted Language Learning
  and Teaching (IJCALLT)}, 5(3):60--74.

\bibitem[{Reusch et~al.(2022)Reusch, Thiele, and
  Lehner}]{reusch2022transformer}
Anja Reusch, Maik Thiele, and Wolfgang Lehner. 2022.
\newblock Transformer-encoder and decoder models for questions on math.
\newblock \emph{Proceedings of the Working Notes of CLEF 2022}, pages 5--8.

\bibitem[{Reyes et~al.(2019)Reyes, Garza, Garrido, De~la Cueva, and
  Ramirez}]{reyes2019methodology}
Roberto Reyes, David Garza, Leonardo Garrido, V{\'\i}ctor De~la Cueva, and
  Jorge Ramirez. 2019.
\newblock Methodology for the implementation of virtual assistants for
  education using google dialogflow.
\newblock In \emph{Advances in Soft Computing: 18th Mexican International
  Conference on Artificial Intelligence, MICAI 2019, Xalapa, Mexico, October
  27--November 2, 2019, Proceedings 18}, pages 440--451. Springer.

\bibitem[{Richey et~al.(2021)Richey, Zhang, Das, Andres-Bray, Scruggs,
  Mogessie, Baker, and McLaren}]{10.1007/978-3-030-78292-4_28}
J.~Elizabeth Richey, Jiayi Zhang, Rohini Das, Juan~Miguel Andres-Bray, Richard
  Scruggs, Michael Mogessie, Ryan~S. Baker, and Bruce~M. McLaren. 2021.
\newblock Gaming and confrustion explain learning advantages for a math digital
  learning game.
\newblock In \emph{Artificial Intelligence in Education}, pages 342--355, Cham.
  Springer International Publishing.

\bibitem[{Rumberg et~al.(2021)Rumberg, Ehlert, L{\"u}dtke, and
  Ostermann}]{rumberg2021age}
Lars Rumberg, Hanna Ehlert, Ulrike L{\"u}dtke, and J{\"o}rn Ostermann. 2021.
\newblock Age-invariant training for end-to-end child speech recognition using
  adversarial multi-task learning.
\newblock In \emph{Interspeech}, pages 3850--3854.

\bibitem[{Sahay et~al.(2019)Sahay, Kumar, Okur, Syed, and
  Nachman}]{KidSpace-SemDial-2019}
Saurav Sahay, Shachi~H. Kumar, Eda Okur, Haroon Syed, and Lama Nachman. 2019.
\newblock \href {http://semdial.org/anthology/Z19-Sahay_semdial_0019.pdf}
  {Modeling intent, dialog policies and response adaptation for goal-oriented
  interactions}.
\newblock In \emph{Proceedings of the 23rd Workshop on the Semantics and
  Pragmatics of Dialogue - Full Papers}, London, United Kingdom. SEMDIAL.

\bibitem[{Sahay et~al.(2021)Sahay, Okur, Hakim, and
  Nachman}]{KS2.0-DaSH-NAACL-2021}
Saurav Sahay, Eda Okur, Nagib Hakim, and Lama Nachman. 2021.
\newblock \href {https://doi.org/10.18653/v1/2021.dash-1.5} {Semi-supervised
  interactive intent labeling}.
\newblock In \emph{Proceedings of the Second Workshop on Data Science with
  Human in the Loop: Language Advances}, pages 31--40, Online. Association for
  Computational Linguistics.

\bibitem[{Sanh et~al.(2019)Sanh, Debut, Chaumond, and
  Wolf}]{sanh2019distilbert}
Victor Sanh, Lysandre Debut, Julien Chaumond, and Thomas Wolf. 2019.
\newblock \href {http://arxiv.org/abs/1910.01108} {Distilbert, a distilled
  version of bert: smaller, faster, cheaper and lighter}.
\newblock In \emph{5th EMC2 Workshop - Energy Efficient Training and Inference
  of Transformer Based Models, 33rd Conference on Neural Information Processing
  Systems (NeurIPS 2019)}.

\bibitem[{Serban et~al.(2018)Serban, Lowe, Henderson, Charlin, and
  Pineau}]{serban2018survey}
Iulian~Vlad Serban, Ryan Lowe, Peter Henderson, Laurent Charlin, and Joelle
  Pineau. 2018.
\newblock A survey of available corpora for building data-driven dialogue
  systems: The journal version.
\newblock \emph{Dialogue \& Discourse}, 9(1):1--49.

\bibitem[{Shen et~al.(2021)Shen, Yamashita, Prihar, Heffernan, Wu, and
  Lee}]{MathBERT-2021}
Jia~Tracy Shen, Michiharu Yamashita, Ethan Prihar, Neil~T. Heffernan, Xintao
  Wu, and Dongwon Lee. 2021.
\newblock \href {http://arxiv.org/abs/2106.07340} {Mathbert: {A} pre-trained
  language model for general {NLP} tasks in mathematics education}.
\newblock \emph{CoRR}, abs/2106.07340.

\bibitem[{Shivakumar and Georgiou(2020)}]{shivakumar2020transfer}
Prashanth~Gurunath Shivakumar and Panayiotis Georgiou. 2020.
\newblock Transfer learning from adult to children for speech recognition:
  Evaluation, analysis and recommendations.
\newblock \emph{Computer speech \& language}, 63:101077.

\bibitem[{Shivakumar et~al.(2014)Shivakumar, Potamianos, Lee, and
  Narayanan}]{shivakumar2014}
Prashanth~Gurunath Shivakumar, Alexandros Potamianos, Sungbok Lee, and
  Shrikanth Narayanan. 2014.
\newblock \href
  {https://www.isca-speech.org/archive_v0/wocci_2014/papers/wc14_015.pdf}
  {Improving speech recognition for children using acoustic adaptation and
  pronunciation modeling}.
\newblock In \emph{Fourth Workshop on Child Computer Interaction (WOCCI 2014)}.

\bibitem[{Shuster et~al.(2022)Shuster, Xu, Komeili, Ju, Smith, Roller, Ung,
  Chen, Arora, Lane et~al.}]{shuster2022blenderbot}
Kurt Shuster, Jing Xu, Mojtaba Komeili, Da~Ju, Eric~Michael Smith, Stephen
  Roller, Megan Ung, Moya Chen, Kushal Arora, Joshua Lane, et~al. 2022.
\newblock Blenderbot 3: a deployed conversational agent that continually learns
  to responsibly engage.
\newblock \emph{arXiv preprint arXiv:2208.03188}.

\bibitem[{Skene et~al.(2022)Skene, O’Farrelly, Byrne, Kirby, Stevens, and
  Ramchandani}]{skene2022can}
Kayleigh Skene, Christine~M O’Farrelly, Elizabeth~M Byrne, Natalie Kirby,
  Eloise~C Stevens, and Paul~G Ramchandani. 2022.
\newblock Can guidance during play enhance children’s learning and
  development in educational contexts? a systematic review and meta-analysis.
\newblock \emph{Child Development}.

\bibitem[{Stemmer et~al.(2017)Stemmer, Georges, Hofer, Rozen, Bauer, Nowicki,
  Bocklet, Colett, Falik, Deisher et~al.}]{stemmer2017speech}
Georg Stemmer, Munir Georges, Joachim Hofer, Piotr Rozen, Josef~G Bauer, Jakub
  Nowicki, Tobias Bocklet, Hannah~R Colett, Ohad Falik, Michael Deisher, et~al.
  2017.
\newblock Speech recognition and understanding on hardware-accelerated dsp.
\newblock In \emph{Interspeech}, pages 2036--2037.

\bibitem[{Stemmer et~al.(2003)Stemmer, Hacker, Steidl, and
  N{\"o}th}]{stemmer2003acoustic}
Georg Stemmer, Christian Hacker, Stefan Steidl, and Elmar N{\"o}th. 2003.
\newblock Acoustic normalization of children's speech.
\newblock In \emph{Eighth European Conference on Speech Communication and
  Technology}.

\bibitem[{Sun et~al.(2021)Sun, Play, Nambiar, and Vidyasagaran}]{sungamifying}
Yueqiu Sun, Tangible Play, Rohitkrishna Nambiar, and Vivek Vidyasagaran. 2021.
\newblock \href {https://mathai4ed.github.io/papers/papers/paper_11.pdf}
  {Gamifying math education using object detection}.
\newblock In \emph{Workshop on Math AI for Education (MATHAI4ED), 35th
  Conference on Neural Information Processing Systems (NeurIPS 2021)}.

\bibitem[{Suresh et~al.(2022{\natexlab{a}})Suresh, Jacobs, Harty, Perkoff,
  Martin, and Sumner}]{suresh-etal-2022-talkmoves}
Abhijit Suresh, Jennifer Jacobs, Charis Harty, Margaret Perkoff, James~H.
  Martin, and Tamara Sumner. 2022{\natexlab{a}}.
\newblock \href {https://aclanthology.org/2022.lrec-1.497} {The {T}alk{M}oves
  dataset: K-12 mathematics lesson transcripts annotated for teacher and
  student discursive moves}.
\newblock In \emph{Proceedings of the Thirteenth Language Resources and
  Evaluation Conference}, pages 4654--4662, Marseille, France. European
  Language Resources Association.

\bibitem[{Suresh et~al.(2022{\natexlab{b}})Suresh, Jacobs, Perkoff, Martin, and
  Sumner}]{suresh-etal-2022-fine}
Abhijit Suresh, Jennifer Jacobs, Margaret Perkoff, James~H. Martin, and Tamara
  Sumner. 2022{\natexlab{b}}.
\newblock \href {https://doi.org/10.18653/v1/2022.bea-1.11} {Fine-tuning
  transformers with additional context to classify discursive moves in
  mathematics classrooms}.
\newblock In \emph{Proceedings of the 17th Workshop on Innovative Use of NLP
  for Building Educational Applications (BEA 2022)}, pages 71--81, Seattle,
  Washington. Association for Computational Linguistics.

\bibitem[{Tack and Piech(2022)}]{tack2022ai}
Ana{\"\i}s Tack and Chris Piech. 2022.
\newblock The ai teacher test: Measuring the pedagogical ability of blender and
  gpt-3 in educational dialogues.
\newblock In \emph{Proceedings of the 15th International Conference on
  Educational Data Mining}, page 522.

\bibitem[{Taghipour and Ng(2016)}]{taghipour-ng-2016-neural}
Kaveh Taghipour and Hwee~Tou Ng. 2016.
\newblock \href {https://doi.org/10.18653/v1/D16-1193} {A neural approach to
  automated essay scoring}.
\newblock In \emph{Proceedings of the 2016 Conference on Empirical Methods in
  Natural Language Processing}, pages 1882--1891, Austin, Texas. Association
  for Computational Linguistics.

\bibitem[{Torpey(2012)}]{torpey2012math}
Elka Torpey. 2012.
\newblock Math at work: Using numbers on the job.
\newblock \emph{Occupational Outlook Quarterly}, 56(3):2--13.

\bibitem[{Tyen et~al.(2022)Tyen, Brenchley, Caines, and
  Buttery}]{tyen-etal-2022-towards}
Gladys Tyen, Mark Brenchley, Andrew Caines, and Paula Buttery. 2022.
\newblock \href {https://doi.org/10.18653/v1/2022.bea-1.28} {Towards an
  open-domain chatbot for language practice}.
\newblock In \emph{Proceedings of the 17th Workshop on Innovative Use of NLP
  for Building Educational Applications (BEA 2022)}, pages 234--249, Seattle,
  Washington. Association for Computational Linguistics.

\bibitem[{Uesato et~al.(2022)Uesato, Kushman, Kumar, Song, Siegel, Wang,
  Creswell, Irving, and Higgins}]{uesatosolving}
Jonathan Uesato, Nate Kushman, Ramana Kumar, H~Francis Song, Noah~Yamamoto
  Siegel, Lisa Wang, Antonia Creswell, Geoffrey Irving, and Irina Higgins.
  2022.
\newblock \href {https://mathai2022.github.io/papers/26.pdf} {Solving math word
  problems with process-based and outcome-based feedback}.
\newblock In \emph{Workshop MATH-AI: Toward Human-Level Mathematical Reasoning,
  36th Conference on Neural Information Processing Systems (NeurIPS 2022), New
  Orleans, Louisiana, USA}.

\bibitem[{Vanzo et~al.(2019)Vanzo, Bastianelli, and
  Lemon}]{vanzo-etal-2019-hierarchical}
Andrea Vanzo, Emanuele Bastianelli, and Oliver Lemon. 2019.
\newblock \href {https://doi.org/10.18653/v1/W19-5931} {Hierarchical multi-task
  natural language understanding for cross-domain conversational {AI}: {HERMIT}
  {NLU}}.
\newblock In \emph{Proceedings of the 20th Annual SIGdial Meeting on Discourse
  and Dialogue}, pages 254--263, Stockholm, Sweden. Association for
  Computational Linguistics.

\bibitem[{Vaswani et~al.(2017)Vaswani, Shazeer, Parmar, Uszkoreit, Jones,
  Gomez, Kaiser, and Polosukhin}]{DBLP:conf/nips/VaswaniSPUJGKP17}
Ashish Vaswani, Noam Shazeer, Niki Parmar, Jakob Uszkoreit, Llion Jones,
  Aidan~N. Gomez, Lukasz Kaiser, and Illia Polosukhin. 2017.
\newblock \href {http://papers.nips.cc/paper/7181-attention-is-all-you-need}
  {Attention is all you need}.
\newblock In \emph{Advances in Neural Information Processing Systems 30: Annual
  Conference on Neural Information Processing Systems 2017, 4-9 December 2017,
  Long Beach, CA, {USA}}, pages 5998--6008.

\bibitem[{Wambsganss et~al.(2020)Wambsganss, Winkler, S{\"o}llner, and
  Leimeister}]{wambsganss2020conversational}
Thiemo Wambsganss, Rainer Winkler, Matthias S{\"o}llner, and Jan~Marco
  Leimeister. 2020.
\newblock A conversational agent to improve response quality in course
  evaluations.
\newblock In \emph{Extended Abstracts of the 2020 CHI conference on human
  factors in computing systems}, pages 1--9.

\bibitem[{Wen et~al.(2018)Wen, Wang, Dong, and Chen}]{wen-2018}
Liyun Wen, Xiaojie Wang, Zhenjiang Dong, and Hong Chen. 2018.
\newblock Jointly modeling intent identification and slot filling with
  contextual and hierarchical information.
\newblock In \emph{Natural Language Processing and Chinese Computing}, pages
  3--15, Cham. Springer International Publishing.

\bibitem[{Winkler et~al.(2020)Winkler, Hobert, Salovaara, S{\"o}llner, and
  Leimeister}]{winkler2020sara}
Rainer Winkler, Sebastian Hobert, Antti Salovaara, Matthias S{\"o}llner, and
  Jan~Marco Leimeister. 2020.
\newblock Sara, the lecturer: Improving learning in online education with a
  scaffolding-based conversational agent.
\newblock In \emph{Proceedings of the 2020 CHI conference on human factors in
  computing systems}, pages 1--14.

\bibitem[{Winkler and S{\"o}llner(2018)}]{alex254848}
Rainer Winkler and Matthias S{\"o}llner. 2018.
\newblock \href {http://www.alexandria.unisg.ch/254848/} {Unleashing the
  potential of chatbots in education: A state-of-the-art analysis}.
\newblock In \emph{Academy of Management Annual Meeting (AOM)}.

\bibitem[{Wollny et~al.(2021)Wollny, Schneider, Di~Mitri, Weidlich, Rittberger,
  and Drachsler}]{wollny2021we}
Sebastian Wollny, Jan Schneider, Daniele Di~Mitri, Joshua Weidlich, Marc
  Rittberger, and Hendrik Drachsler. 2021.
\newblock Are we there yet?-a systematic literature review on chatbots in
  education.
\newblock \emph{Frontiers in artificial intelligence}, 4:654924.

\bibitem[{Wu et~al.(2019)Wu, Garc{\'\i}a-Perera, Povey, and
  Khudanpur}]{wu2019advances}
Fei Wu, Leibny~Paola Garc{\'\i}a-Perera, Daniel Povey, and Sanjeev Khudanpur.
  2019.
\newblock Advances in automatic speech recognition for child speech using
  factored time delay neural network.
\newblock In \emph{Interspeech}, pages 1--5.

\bibitem[{Wu et~al.(2018)Wu, Fisch, Chopra, Adams, and
  Weston}]{wu2018starspace}
Ledell Wu, Adam Fisch, Sumit Chopra, Keith Adams, and Antoine Bordes~Jason
  Weston. 2018.
\newblock Starspace: Embed all the things!
\newblock In \emph{Proceedings of the Thirty-Second AAAI Conference on
  Artificial Intelligence and Thirtieth Innovative Applications of Artificial
  Intelligence Conference and Eighth AAAI Symposium on Educational Advances in
  Artificial Intelligence}, AAAI'18/IAAI'18/EAAI'18. AAAI Press.

\bibitem[{Yang et~al.(2022)Yang, Qin, Chen, Lin, and
  Liang}]{yang-etal-2022-logicsolver}
Zhicheng Yang, Jinghui Qin, Jiaqi Chen, Liang Lin, and Xiaodan Liang. 2022.
\newblock \href {https://aclanthology.org/2022.findings-emnlp.1}
  {{L}ogic{S}olver: Towards interpretable math word problem solving with
  logical prompt-enhanced learning}.
\newblock In \emph{Findings of the Association for Computational Linguistics:
  EMNLP 2022}, pages 1--13, Abu Dhabi, United Arab Emirates. Association for
  Computational Linguistics.

\bibitem[{Yang et~al.(2019)Yang, Dai, Yang, Carbonell, Salakhutdinov, and
  Le}]{yang2019xlnet}
Zhilin Yang, Zihang Dai, Yiming Yang, Jaime Carbonell, Russ~R Salakhutdinov,
  and Quoc~V Le. 2019.
\newblock \href
  {https://proceedings.neurips.cc/paper_files/paper/2019/file/dc6a7e655d7e5840e66733e9ee67cc69-Paper.pdf}
  {Xlnet: Generalized autoregressive pretraining for language understanding}.
\newblock In \emph{Advances in Neural Information Processing Systems},
  volume~32. Curran Associates, Inc.

\bibitem[{Yeung and Alwan(2018)}]{yeung2018difficulties}
Gary Yeung and Abeer Alwan. 2018.
\newblock On the difficulties of automatic speech recognition for
  kindergarten-aged children.
\newblock \emph{Interspeech 2018}.

\bibitem[{Yeung et~al.(2021)Yeung, Fan, and Alwan}]{yeung2021fundamental}
Gary Yeung, Ruchao Fan, and Abeer Alwan. 2021.
\newblock Fundamental frequency feature normalization and data augmentation for
  child speech recognition.
\newblock In \emph{ICASSP 2021-2021 IEEE International Conference on Acoustics,
  Speech and Signal Processing (ICASSP)}, pages 6993--6997. IEEE.

\bibitem[{Zhai et~al.(2021)Zhai, Chu, Chai, Jong, Istenic, Spector, Liu, Yuan,
  and Li}]{zhai2021review}
Xuesong Zhai, Xiaoyan Chu, Ching~Sing Chai, Morris Siu~Yung Jong, Andreja
  Istenic, Michael Spector, Jia-Bao Liu, Jing Yuan, and Yan Li. 2021.
\newblock A review of artificial intelligence ({AI}) in education from 2010 to
  2020.
\newblock \emph{Complexity}, 2021.

\bibitem[{Zhang and Wang(2016)}]{zhang-2016}
Xiaodong Zhang and Houfeng Wang. 2016.
\newblock \href {http://dl.acm.org/citation.cfm?id=3060832.3061040} {A joint
  model of intent determination and slot filling for spoken language
  understanding}.
\newblock In \emph{Proceedings of the Twenty-Fifth International Joint
  Conference on Artificial Intelligence}, IJCAI'16, pages 2993--2999. AAAI
  Press.

\bibitem[{Zhang et~al.(2020)Zhang, Sun, Galley, Chen, Brockett, Gao, Gao, Liu,
  and Dolan}]{zhang-etal-2020-dialogpt}
Yizhe Zhang, Siqi Sun, Michel Galley, Yen-Chun Chen, Chris Brockett, Xiang Gao,
  Jianfeng Gao, Jingjing Liu, and Bill Dolan. 2020.
\newblock \href {https://doi.org/10.18653/v1/2020.acl-demos.30} {{DIALOGPT} :
  Large-scale generative pre-training for conversational response generation}.
\newblock In \emph{Proceedings of the 58th Annual Meeting of the Association
  for Computational Linguistics: System Demonstrations}, pages 270--278,
  Online. Association for Computational Linguistics.

\bibitem[{Zhang et~al.(2022)Zhang, Xu, Wang, Yao, Ritchie, Wu, Yu, Wang, and
  Li}]{zhang2022storybuddy}
Zheng Zhang, Ying Xu, Yanhao Wang, Bingsheng Yao, Daniel Ritchie, Tongshuang
  Wu, Mo~Yu, Dakuo Wang, and Toby Jia-Jun Li. 2022.
\newblock Storybuddy: A human-ai collaborative chatbot for parent-child
  interactive storytelling with flexible parental involvement.
\newblock In \emph{Proceedings of the 2022 CHI Conference on Human Factors in
  Computing Systems}, pages 1--21.

\end{thebibliography}
\bibliographystyle{acl_natbib}


\appendix

\section{Appendix: Additional Error Analysis}
\label{sec:appendix}

Please refer to Table~\ref{asr-errors} for additional error analysis on ASR output from our home deployment data. Here, we compare manually transcribed utterances (i.e., human transcripts) with the speech recognition output (i.e., raw ASR transcripts) using five different ASR models that we investigated in this study. These ASR errors demonstrate the challenges faced in the speech recognition model performances on kids' speech, which potentially would be propagated into the remaining modules in the conventional task-oriented dialogue pipeline. 

We may attribute various factors to these speech recognition errors, often related to our deployment data characteristics. Incidental voices and phrases constitute a good chunk of the overall home deployment data, along with very short utterances to be recognized (e.g., stating names, colors, types of flowers, numbers, and binary answers with one-or-two words), plus the remaining known challenges present with recognizing kids' speech in noisy real-world environments.

\begin{table*}[!ht]
  \centering
  \resizebox{\textwidth}{!}{
  \begin{tabular}{llllll}
    \toprule
    \textbf{Human Transcript} & \textbf{Rockhopper} & \textbf{Google Cloud} & \textbf{Whisper-base} & \textbf{Whisper-small} & \textbf{Whisper-medium}\\
    \midrule
    \midrule
    Atticus. & - & - & Yeah, that's cute. & I have a kiss. & Now I have to kiss. \\
    \midrule
    I am Genevieve. & i'm twenty-two & I'm going to be & I'm Kennedy. & I'm Genevieve. & I'm Genevieve. \\
    \midrule
    Red. & rab & - & Ralph. & Red. & Red. \\
    \midrule
    Blue. & lil & blue & Blair. & Blue. & Blue. \\
    \midrule
    Yes, & laughs & yes & Yes? & Yes? & Yes? \\
    \midrule
    Roses. & it is & roses & Okay. & Okay & focus \\
    \midrule
    Zero. & you know & no & No. & No, no. & No. \\
    \midrule
    four. & you swore & - & forward. & Over. & Over. \\
    \midrule
    five. & - & bye & Bye. & Bye. & Bye. \\
    \midrule
    eight & all & - & Thank you. & Bye. & Oh \\
    \midrule
    forty eight & wall e & 48 & 48 & 48 & 48 \\
    \midrule
    forty nine & already & 49 & 49 & 49 & 49 \\
    \midrule
    fifty one & if you want & 51 & 51 & 51 & 51 \\
    \midrule
    seventy four & stopping before & 74 & 74 & 74 & 74 \\
    \midrule
    Maybe tomorrow. & novarro & tomorrow & I need some water, & I'm going to leave & I'm leaving tomorrow. \\
     & & & though. & it tomorrow. & \\
    \midrule
    Flowers, flowers in & lean forward & Greenhouse & In forward, in & I think forward, & In the green house. \\
    the greenhouse? &  phelps hours & & forward, in the & both flowers and & \\
     & than we & & green house. & the greenhouse. & \\
    \midrule
    There are seventeen, & seventeen & 17 + 17 - 27 & There are 17 and & There are 17 and & What is the maximum \\
    and seventeen minus & seventeen & & 17 minus 10 & 17 minus 10 & number of children in the \\
    ten equals seven. & rooms & & equals 7. & equals 7. & world? Um... There are 17 \\
     & & & & & and 17 minus 10 equals 7. \\
    \bottomrule
  \end{tabular}
  }
  \caption{ASR Error Samples from Kid Space Home Deployment Data}
  \label{asr-errors}
\end{table*}

\end{document}